\documentclass[11pt,thmsa]{article}%
\usepackage{amssymb}
\usepackage{amsmath}
\usepackage{setspace}
\usepackage{amsfonts}
\usepackage{graphicx}
\usepackage[round,sort,authoryear]{natbib}
\usepackage{mathtools}
\usepackage[colorlinks=true,linkcolor=blue, citecolor=blue, urlcolor = blue]{hyperref}
\usepackage{multirow}

\providecommand{\U}[1]{\protect\rule{.1in}{.1in}}
\newtheorem{theorem}{Theorem}

\newtheorem{algorithm}{Algorithm}

\newtheorem{corollary}{Corollary}

\newtheorem{example}{Example}

\newtheorem{proposition}{Proposition}
\newtheorem{remark}{Remark}

\newtheorem{assumption}{Assumption}

\newenvironment{proof}[1][Proof]{\noindent\textbf{#1.} }{\ \rule{0.5em}{0.5em}}
\advance\oddsidemargin by -0.35 in
\advance\evensidemargin by -0.35 in
\newcommand\norm[1]{\left\lVert#1\right\rVert}

\begin{document}

\title{ \textbf{Optimal Portfolio Choice and Stock Centrality for Tail Risk Events}  }

\author{ \textbf{Christis Katsouris}\thanks{Department of Economics, School of Social Sciences, University of Southampton, Southampton, SO17 1BJ, United Kingdom. \textit{E-mail}: \textcolor{blue}{C.Katsouris@soton.ac.uk}. I am grateful to my advisor Prof. Jose Olmo for providing constructive feedback that improved earlier versions of the paper. I also would like to thank Prof. Peter W. Smith for helpful discussion. Financial support from the VC PhD studentship of the University of Southampton is gratefully acknowledged. The author also acknowledge the use of the IRIDIS HPC Facility and associated support services at the University of Southampton in the completion of this work.} \\ \textbf{University of Southampton}   }

\maketitle

\begin{abstract}
We propose a novel risk matrix to characterize the optimal portfolio choice of an investor with tail concerns. The diagonal of the matrix contains the Value-at-Risk of each asset in the portfolio and the off-diagonal the pairwise $\Delta$CoVaR measures reflecting tail connections between assets. First, we derive the conditions under which the associated quadratic risk function has a closed-form solution. Second, we examine the relationship between portfolio risk and eigenvector centrality. Third, we show that portfolio risk is not necessarily increasing with respect to stock centrality. Forth, we demonstrate under certain conditions that asset centrality increases the optimal weight allocation of the asset to the portfolio. Overall, our empirical study indicates that a network topology which exhibits low connectivity is outperformed by high connectivity based on a Sharpe ratio test. 
\\

\end{abstract}

%
\newpage 

\section{Introduction}

The traditional portfolio allocation problem has been extensively studied in the literature the past decades, providing a way of finding efficient allocations among expected return and risk. However, considering the traditional trade-off as a stand alone dilemma is not a sufficient decision mechanism as the 2008 global financial crisis revealed. In this paper, building on the optimal portfolio optimization problem of \cite{markowitz1956optimization} (see, also \cite{sharpe1964capital}) we propose a novel framework within which the network topology is incorporated into a graph-based portfolio selection mechanism. Therefore, our main goal in this paper is the introduction of a novel measure of centrality when the objective function characterizing the portfolio weights is a tail risk measure instead of the traditional variance and covariance. 

Following our main objective, we aim to address two main research questions. First, what is the mechanism of optimally allocating assets based on a graph representation of the portfolio? Specifically, we propose conditions under which a network based risk matrix permits to construct an equivalent quadratic loss function that corresponds to the variance-covariance matrix. Second, given the construction of our tail-based risk matrix, what is the role of stock centrality in the optimal weight allocation and how does it affect estimated portfolio risk based on the network tail risk matrix?  
 
To answer the aforementioned research questions and provide theoretical and empirical evidence related to these aspects, we explore the relationship between stock centrality and portfolio risk with focus on eigenvector centrality (see, \cite{Bonacich72}). Specifically, we investigate the role of stock centrality on portfolio risk estimated based on our novel risk matrix which is designed to capture tail events. Moreover, the proposed risk matrix provides a simple way to construct an adjacency matrix which gives a mathematical representation of the network topology. A related study is presented by \cite{peralta2016network} who examine the role of eigenvector centrality on the optimal portfolio vector of weights based on the correlation matrix, however our approach and conclusions provide a different perspective. 

Our framework considers the measurement of financial connectedness in the estimation of optimal exposures to stocks for an investment portfolio. More specifically, by considering the quadratic loss function based on the network topology we can decompose the elements generated by the network tail risk matrix shedding light on how the diversification effect occurs within our setting. The associated quadratic risk function characterizes the optimal portfolio and consists of a sum of idiosyncratic risks, defined as the VaR risk measures located in the main diagonal of the risk matrix, and the pairwise connections between the assets in the tails, defined as the $\Delta \text{CoVaR}$ risk measures, located in the off-diagonal of the risk matrix. 

Some important results emerge from our theoretical derivations such that the portfolio risk does not necessarily increase with asset centrality (see, also \cite{Olmo2021AssetCentr}. Then, by imposing further optimization constraints such as portfolios only taking long positions on the assets and centrality measures only taking positive values, then we can show that portfolio risk increases with asset centrality. To provide theoretical evidence we examine the relationship between the centrality of an asset and the magnitude of its associated optimal portfolio allocation. We find that under general conditions the higher the centrality of an asset, as explained by eigenvector centrality, the larger its optimal allocation to the investment portfolio. 

This result clearly contrasts the findings presented by  \cite{pozzi2013spread} and \cite{peralta2016network}. Specifically, the particular studies establish that optimal portfolio strategies require to overweight stocks with low eigenvector centrality and underweight stocks with high centrality in the network. Based on the estimation frameworks of these studies, optimal investors attempt to benefit from diversification by avoiding the allocation of wealth towards assets that are central in the correlation-based network. On the other hand, based on the proposed estimation framework of this paper, using a more general network driven tail risk matrix we find evidence of the opposite result to hold. In particular, we find theoretical and empirical evidence that is optimal to overweight central assets when investors allocate wealth using a quadratic loss function characterized by our tail risk matrix. 

To align these seemingly contradictory findings, regarding the relationship between portfolio risk and asset centrality we illustrate that these are not specific  to our novel risk matrix. To demonstrate this, we show the role of asset centrality in a standard minimum variance framework in which the risk matrix represents the conventional variance-covariance matrix. In this traditional setting the main results in the paper establishing the relationship between portfolio risk, optimal asset allocation and asset centrality hold, under general regulatory conditions which can be directly applied to the proposed graph based procedure. 

The contributions of this paper are as follows. We propose a novel risk matrix given by the VaR and $\Delta \text{CoVaR}$ of the assets in a portfolio. This risk matrix aims to capture tail events given by idiosyncratic tail risk and connectivity between assets in the tails. In this setting, investors minimize a loss function similar to the minimum variance portfolio introduced by \cite{Markowitz52} in which the variance-covariance matrix is replaced by the tail-based risk matrix. We derive the conditions under which the associated quadratic form has a single solution that is interpreted as the optimal portfolio allocation of an investor with portfolio tail concerns. To the best of our knowledge the examination of these components within a unified framework consist of a novel contribution to the literature. For example, \cite{Olmo2021AssetCentr} proves that the relationship between eigenvector centrality and optimal portfolio weight is not monotonic as previously thought in the literature (see \cite{peralta2016network}). 

The rest of the paper is organized as follows. Section \ref{SectionI} introduces the optimal portfolio allocation framework under tail risk events including network tail dependencies. Section \ref{SectionII} presents the main result of the paper, namely, the existence of a positive relationship between asset centrality and its optimal asset allocation. Section \ref{SectionIII} studies the role of asset centrality across portfolios of different size and explores the effect of the centrality of an asset on the centrality of the remaining assets. Section \ref{SectionIV} presents the numerical implementations of our framework. Section \ref{SectionV} concludes. Technical proofs can be found in the Appendix.

\newpage

\paragraph{Related Literature} A large steam of literature propose various statistical methods for the measurement of financial connectedness and systemic risk. For instance, related studies include \cite{hong2009granger}, \cite{billio2012econometric}, \cite{diebold2012better, diebold2014network}, \cite{corsi2018measuring} who consider the construction of Granger causality networks as well as \cite{hardle2016tenet} who propose a nonlinear network based tail risk matrix. Moreover, \cite{anufriev2015connecting} and \cite{hale2019monitoring} consider the effect of the network topology in the monitoring of systemic risk in the banking sector (see, also \cite{huber2018disentangling}). According to \cite{mitchener2019network}, the investigation of the causality of shock propagation, amplification of systemic risk and potentially the magnitude of economic downturns within a network remains an open economic question. \cite{calvo2021granger} propose a novel framework to identify Granger causality using neural network models which can be utilized for explaining the propagation of systemic risk in high-dimensional settings.  
    
Firstly, quantile measures are commonly used as risk measures in optimal portfolio allocation problems, see among others, \cite{DuffiePan97},  \cite{Jorion07} and \cite{white2015var} for a comprehensive review of VaR models in parametric settings. In an optimal asset allocation context, VaR quantiles usually act as an optimization constraint rather than as a target variable to be optimized. The mean-risk models discussed by \cite{Fishburn77} can be considered as an extension of the standard mean-variance formulations of Markowitz. Further relevant literature includes \cite{BasakShapiro01}, \cite{KrokhmalUryasevPalmquist01}, \cite{CampbellHoismanKoedijk01}. In addition, the studies of \cite{WuXiao02}, \cite{BassettKoenkerKordas:04} as well as the paper of \cite{EngleManganelli:04} and \cite{IbragimovWalden:07}, among others, shed an interesting light on the properties of VaR-optimal portfolios while acknowledging considerable computational difficulties (see, \cite{GaivoronskiPflug05}).

Secondly, a commonly used methodology for the analysis of mean-variance portfolios employs the multivariate dynamic conditional correlation GARCH model proposed by \cite{engle2002dynamic} which is based on the well documented GARCH(p,q) model of \cite{bollerslev1986generalized}. The particular class of dynamic conditional volatility matrices provide a way for constructing robust dynamic optimal portfolios using an out-of-sample forecasting scheme. Furthermore, in high dimensional settings such dynamic covariance matrices can suffer from the curse of dimensionality which can affect the estimation accuracy of the portfolio allocation. For instance, \cite{abadir2014design} propose a sample splitting procedure to combine estimates for the eigenvalues and the eigenvectors of the covariance matrix. Specifically, in the case of high dimensional portfolios, to overcome the related accuracy issues \cite{ledoit2020analytical} and \cite{engle2019large} propose shrinkage and non-linear estimators of the dynamic covariance matrix (see also \cite{avella2018robust}). Lastly, it is worth mentioning in this stream of literature \cite{bollerslev2020realized} who propose a novel risk matrix based on semi-covariances.  

Thirdly, centrality measures are commonly used to provide a statistical representation of the level of connectedness of a node and to access spillover effects within the network. \cite{ballester2006s} propose a framework in which node centrality can employed to examine equilibrium relations in a graph. Furthermore, the aspects of financial contagion have been examined in the literature as an equilibrium phenomenon, such as in  \cite{allen2000financial}. Additionally, \cite{duffie2009frailty} propose modelling correlated defaults in financial markets with correlated hazard rates based on a cross-sectional dataset of firms. 

Lastly, a framework for the role of correlation risk in the optimal portfolio choice problem is presented by \cite{buraschi2010correlation} (see, \cite{moreira2017volatility}). The particular investment interdependence is reported in the literature and goes beyond portfolio optimization (see, \cite{lamont1997cash} and \cite{stein1997internal}). Nevertheless,  if one aims to understand the structure of the market and use it for, say, financial stability analysis or optimal policy design, then our proposed tail risk matrix provides an analytically tractable way of examining the effect of asset centrality and portfolio risk. Our approach considers single index specification models to model tail risk, that is, VaR and CoVaR among financial institutions, focusing this way on the relation between optimal portfolio allocation problem and asset centrality via the proposed novel risk matrix. Therefore,  our framework which focus on the network tail-risk matrix rather that optimizing portfolio performance measures when those are expressed in terms of systemic risk measures as in \cite{bergk2021portfolio} and \cite{LinOlmo2021Portfolio}. 

Throughout the paper, we denote with $\frac{ \partial  }{ \partial x} f(x)$ and $\frac{ \partial^2  }{ \partial x^2} f(x)$ the first and second partial derivatives of a continuous function for some $x$ within the support of the function. Moreover, $\left\| \ . \ \right\|_2$ denotes the Euclidean norm, we denote with $\mathbf{1} = \left( 1,1,..., 1 \right)^{\prime}$ the unit vector with length $N$ and $\mathbf{I}_{N}$ to be $N \times N$ identity matrix. Furthermore, for a real-valued matrix $\mathbf{A} = \left( A_{ij} \right)$ such that $\mathbf{A} \in \mathbb{R}^{N \times N}$, we denote with $\mathbf{A}^{-1}$ its inverse matrix, and define the spectral norm to be $\norm{ \mathbf{A} }_2 = \left[ \lambda_{\text{max}} \left( \mathbf{A}^{\prime} \mathbf{A} \right) \right]^{1/2}$, where $\lambda_{\text{max}}$ and $\lambda_{\text{min}}$
the largest and smallest eigenvalue respectively, the matrix $\ell_1$ norm to be $\norm{ \mathbf{A} }_1 = \text{max}_j \sum_i \left| A_{ij} \right|$ and the entrywise $\ell_{ \infty }$ norm to be $\left| \mathbf{A} \right|_{\infty } = \text{max}_{i,j} \left| A_{ij} \right|$. 

\section{Optimal Portfolio Choice} 
\label{SectionI}

We begin with the traditional optimal portfolio allocation problem. We extend this framework by proposing a novel risk matrix that captures tail events. We present regularity conditions which ensure the existence of a closed-form solution for the associated quadratic risk function. 

\subsection{Minimum-Variance Portfolio Optimization}
\label{SectionI.A}

The Markowitz portfolio optimization theory considers the minimization of the portfolio variance without any further restriction on the corresponding expected return. More formally, we denote with $\mathbf{R}_t = ( R_{1,t} ,..., R_{N,t} )^{\prime}$ the $N-$dimensional vector of stock returns at time $t$,  assumed to be a set of strictly stationary time series. We denote with $\text{\boldmath$\mu$} := \mathbb{E} \left( \mathbf{R}_t \right) = \left( \mathbb{E} \left( R_{1,t} \right) ,..., \mathbb{E} \left( R_{N,t} \right) \right)^{\prime}$ where $\text{\boldmath$\mu$} \in \mathbb{R}^{N \times 1}$ the vector of expected returns at time $t$ and the corresponding variance-covariance matrix defined as $\mathbf{\Sigma} := \text{Cov}\left( \mathbf{R}_t \right) = \mathbb{E} \left( \mathbf{R}_t \mathbf{R}_t^{\prime} \right) - \text{\boldmath$\mu$} \text{\boldmath$\mu$}^{\prime}$.

\newpage 

Let $r_t^{p} = \mathbf{w}_i^{\prime} \mathbf{R}_t$ denote the return on a portfolio of $N$ assets, with $\mathbf{w}_i= \left( w_{1},...,w_{N} \right)^{\prime}$ the vector of portfolio weights representing the financial position of the investor as a weight allocation\footnote{Note that we use the notation,$\mathbf{w} \in \mathbb{R}^{N \times 1}$ to represent the vector of unknown weights and $\mathbf{w}^{*} \in \mathbb{R}^{N \times 1}$ to represent the optimal vector of weights estimated via the optimization problem. For the empirical application of the paper, we assume that the vector of optimal weights is consistently estimated by $\widehat{\mathbf{w}}^{*} \in \mathbb{R}^{N \times 1}$ which implies that since we are using an optimization routine to ensure convergence within the unit ball, in the \textit{L}$_1$-norm we have that $||  \widehat{\mathbf{w}}^{*} -  \mathbf{w}^{*}||_1 = o_p(1)$. } of assets to the portfolio. A common practise in optimal portfolio choice problems is to use the sample moments (i.e., sample mean and covariance, e.g., see \cite{fan2015risks}). The usual sample estimates are given by $\widehat{\text{\boldmath$\mu$}}$ and $\widehat{\mathbf{\Sigma}}$ respectively. 
\begin{align}
\widehat{\text{\boldmath$\mu$}} &:= \bigg[ \frac{1}{T} \sum_{t=1}^T R_{1,t} ,...,   \frac{1}{T} \sum_{t=1}^T R_{N,t} \bigg] 
\\
\widehat{\mathbf{\Sigma}} &:= \frac{1}{T} \sum_{t=1}^T ( \mathbf{R}_t - \bar{\mathbf{R}}) ( \mathbf{R}_t - \bar{\mathbf{R}})^{\prime}, \ \text{where} \ \ \bar{\mathbf{R}} = \frac{1}{T} \sum_{t=1}^T \mathbf{R}_t
\end{align} 
where $\widehat{\mathbf{\Sigma}}:= \left( \hat{\sigma}_{ij} \right)_{N \times N}$ is the sample covariance matrix. Then, the variance of this portfolio is defined as $V \left( r_t^{p} \right) = \mathbf{w}^{\prime} \mathbf{\Sigma} \mathbf{w}$. Therefore, we can now introduce the Markowitz's minimum variance optimization problem which is expressed as below
\begin{align}
\label{the problem}
\underset{\mathbf{w} \in \mathbb{R}^N }{\text{arg min}} \ \big\{  \mathbf{w}^{\prime} \mathbf{\Sigma} \mathbf{w} \big\} \ \text{subject to }  \  \mathbf{w}^{\prime} \mathbf{1}= 1.
\end{align}
The first order conditions to the minimization problem yield the following vector of weights
\begin{align}
\label{minv}
\mathbf{w}^{*} = \frac{\mathbf{\Sigma}^{-1} \mathbf{1}}{\mathbf{1}^{\prime} \mathbf{\Sigma}^{-1} \mathbf{1} }.
\end{align}
Moreover, \cite{Markowitz52} considers the extension of this portfolio allocation problem to a mean-variance efficient frontier. Let $R_{f,t}$ be the risk-free return such that the excess return of the assets in the portfolio is now $R_{t}^e = R_{i,t} - R_{f,t}$ and the vector of expected excess returns is $\boldsymbol{\mu}^e := \mathbb{E} \left( \mathbf{R}_t - R_{f,t} \right)$. The optimal mean-variance portfolio occurs by minimizing the portfolio variance for a given level of expected excess return $\bar{R}$. 

\subsection{Optimal Portfolio Allocation with a Graph-based Tail Risk Matrix}
\label{SectionI.B}

In this section, we refer to centrality as the eigenvector centrality and thus distinguish it explicitly from different types of centrality. The aim is to introduce the framework in which we can explore the optimal asset allocation problem when investors penalize tail risk measures. To do this, we focus on the VaR and $\Delta \text{CoVaR}$ tail measures and propose a novel network based risk matrix which incorporates these risk measures. Our starting point in this direction is to consider a more general structure than the traditional variance-covariance matrix as the risk matrix used in portfolio optimization problems. 

\newpage 

To do this, we replace the variance of an asset return by its Value-at-Risk (VaR) and the cross-correlation measures between asset returns by the Delta Conditional VaR, denoted with $\Delta \text{CoVaR}$, as proposed by \cite{adrian2011covar} to model systemic risk. 

The Value-at-Risk (VaR) of asset $i$ with coverage probability $\tau$ is defined by the condition
\begin{align}
\mathbb{P} \bigg(  R_{i,t+1} \leq \text{VaR}_{i,t} ( \tau ) \ \big| \ \mathcal{F}_t \bigg) = \tau \ \ \ \text{with} \ \ \ \tau \in (0,1),   
\end{align} 
where $\mathcal{F}_t$ denotes the sigma-algebra containing all the information relevant at time $t$. This quantity captures idiosyncratic tail risk, which implies that $\text{VaR}_{i,t} ( \tau )$ can be interpreted as the quantile $\tau$ of the underline conditional distribution of the return $R_{i,t+1}$ given the information available at time $t$. Then, the Conditional Value at Risk (CoVaR) is defined as
\begin{align}
\mathbb{P} \bigg( R_{i,t+1} \leq \text{CoVaR}_{i| \mathcal{C}( R_{j,t+1})} ( \tau ) \ \big| \ \mathcal{F}_t \bigg) = \tau. 
\end{align}
Various authors have proposed suitable expressions for the conditioning set $\mathcal{C} \left( R_{j,t+1} \right)$, which can be used to define a distress event for firm $j$. For instance, according to \cite{adrian2011covar} such a financial distress event is interpreted as the institution's loss which is  equal to the firm's VaR, capturing this way conditional tail risk. In the original definition of CoVaR the return $R_{i,t}$ is replaced by a market return $R_{m,t}$ such that the CoVaR measure captures systemic risk. In what follows, we use the notation $\text{VaR}_{i,t}$ and $\text{CoVaR}_{(i,j),t}$ while omitting the quantile $\tau$ for the sake of notation brevity. 

Then, the CoVaR is a quantile risk measure defined as below
\begin{align}
\mathbb{P} \bigg( R_{i,t+1} \leq \text{CoVaR}_{(i,j),t} ( \tau ) \ \big| \ R_{j,t+1} = \text{VaR}_{j,t} ( \tau )  , \mathcal{F}_t \bigg) = \tau. 
\end{align}
Therefore, the Delta Conditional VaR for a coverage probability $\tau \in (0,1)$ is defined as
\begin{equation}
\Delta \text{CoVaR}_{(i,j),t} ( \tau ) = \text{CoVaR}_{(i,j),t} ( \tau ) - \text{VaR}_{i,t} (\tau).
\end{equation}
Intuitively, the absence of tail dependence between the assets entails the condition that $\Delta \text{CoVaR}_{(i,j),t} ( \tau ) = 0$. Moreover, in line with the literature on risk management and tail portfolio risk, we restrict the quantile tail set into a small region of the underline  distribution, such that $\tau \in ( 0, \epsilon ]$, where $\epsilon < 0.15$. In this region the $\text{VaR}_{i,t} ( \tau) $ and $\text{CoVaR}_{(i,j),t} ( \tau)$ quantile measures are negative since the distribution of asset returns is centred at zero. In particular, these risk measures can be positive or negative, thus, a negative value of $\Delta \text{CoVaR}_{(i,j),t} ( \tau )$ reflects the presence of positive spillovers from asset $j$ to asset $i$ in the tail of the distribution implying a shift on the $\text{VaR}_{i,t}$ quantile measure to the left of the distribution when the return on asset $j$ is at its $\text{VaR}_{j,t}( \tau )$ level. Similarly, a positive value of $\Delta \text{CoVaR}_{(i,j),t} ( \tau )$ reflects negative spillovers from asset $j$ to asset $i$ in the tail of the distribution implying a shift on the $\text{VaR}_{i,t}$ quantile measure to the right of the distribution. 

\newpage 

Throughout the paper we consider the corresponding quantities $-\text{VaR}$, $-\text{CoVaR}$ as the relevant risk measures which appear in the proposed risk matrix. Thus, under the parameter restriction $\epsilon < 0.15$, it holds that\footnote{This property holds due to the construction of these risk measures based on distribution quantiles, such as $q_{\tau}^{+} (X) = inf \left\{ t \in \mathbb{R}| \mathbb{P} \left( X \leq t \right) > \tau \right\}$, where $\tau \in (0,1)$. } $0<-\text{VaR}, -\text{CoVaR}<1$. We proceed by introducing the portfolio choice problem under tail risk events. Thus, we construct a portfolio that is analytically tractable and minimizes the dependence between the assets under distress events. 

The main motivation behind the specification of a risk matrix with tail risk measures, is to examine how the investor's loss function is affected by co-movements in the tails of the underline stock return distributions. As a second objective, we aim to propose an algorithm to eliminate sequentially the stocks from the portfolio to reduce risk given a certain level of returns. Therefore, in contrast to the mean-variance framework, minimization of this objective function does not correspond to minimization of second or higher order moments of the underline distribution of portfolio returns. Our proposed risk matrix is a quadratic form that minimizes the contributions of the idiosyncratic quantile tail risk measures given by the VaR measures and the contributions of the tail dependencies between pairs of assets captured by the $\Delta \text{CoVaR}$ measures which represent the effects of interactions within a financial network.  

In this context, the investor's minimization problem \eqref{minv} at a given time $t$ is replaced by 
\begin{align}
\label{minVaR}
\mathbf{w}^{*} =  \underset{\mathbf{w} \in \mathbb{R}^N }{\text{arg min}} \ \big\{  \mathbf{w}^{\prime} \mathbf{\Gamma}_t  \mathbf{w} \big\} \ \text{subject to }  \  \mathbf{w}^{\prime} \mathbf{1}= 1,
\end{align}
\begin{align} 
\label{gammamat}
\text{ such that } \ \ \mathbf{\Gamma}_t  
=  
\begin{bmatrix}
    \gamma_{1,t}   & \gamma_{(1,2),t}  & \dots \gamma_{(1,N),t} \\
    \vdots       &  \vdots        & \vdots             \\
    \vdots       &  \ddots        & \vdots              \\
    \gamma_{(N,1),t} &  \dots         & \gamma_{N,t}         \\
\end{bmatrix}
\end{align}
\begin{align}
\text{with} \ \gamma_{i,t}= \text{VaR}^{+}_{i,t} \ \text{and} \ \gamma_{(i,j),t}= (\text{VaR}^{+}_{i,t} \text{VaR}^{+}_{j,t})^{1/2} \Delta \text{CoVaR}_{(i,j),t}, \ \ \forall \ i,j \in \left\{1,...,N \right\}
\end{align}
The asymmetry in spillover effects in the tails of the distributions of stocks $i$ and $j$ imply that the risk matrix $\mathbf{\Gamma}_t \in \mathbb{R}^{ N \times N }$ is a non-symmetric square matrix. To apply standard results in portfolio choice problems and to be able to establish a parallelism with the conventional Markowitz's setting, we transform the minimization problem \eqref{minVaR} into an optimization problem characterized by the symmetric counterpart of $\mathbf{\Gamma}_t$, that is denoted as $\widetilde{ \mathbf{ \Gamma} }_t$ hereafter. The symmetrization of the off-diagonal elements of the original matrix $\mathbf{\Gamma}$, guarantees the positive definitiness of the $\widetilde{ \mathbf{\Gamma} }$ matrix. In particular, the symmetization of our proposed risk matrix provides a simple way for the estimation of the concentration matrix avoiding methods such as the use of the adaptive thresholding estimator (see, \cite{avella2018robust}) which is commonly used in estimation of high dimensional covariance matrices. 

\newpage 

Our proposed risk matrix is a regression based matrix with tail forecasts under parametric assumptions. More specifically, the $\left( i, j \right)$th row of $\widetilde{ \mathbf{\Gamma} }$ are estimated via the tail risk measures of VaR$^{+}$ and CoVaR$^{+}$ based on quantile regression specifications of single index models. The specific methodology requires parametric assumptions. For instance, in the case of covariance matrices, other distribution-free techniques are proposed in the literature, see for instance the study of \cite{browne1984asymptotically}. Furthermore, our approach considers tail dependence using conditional quantiles and therefore captures tail risk which are not exchangeable and have causal effects which are not symmetric (see, \cite{bernard2015conditional} and \cite{zhang2008modelling}).  For this reason, within the framework of optimal asset allocation, the risk matrix has to be symmetrized. Therefore, we transform the asymmetric bilinear form
$ \mathcal{Q} (\mathbf{\Gamma}_t,\mathbf{w})=\mathbf{w}^{\prime} \mathbf{\Gamma}_t \mathbf{w}$, into a symmetric bilinear form (see, \cite{csaki1970concise}) defined as following
\begin{align} 
\label{matrix.new.specification}
\mathcal{Q}( \widetilde{ \mathbf{\Gamma} }_t,\mathbf{w})= \mathbf{w}^{\prime} \widetilde{ \mathbf{\Gamma} }_t \mathbf{w}, 
\end{align}
where 
\begin{align}
\label{matrix.new.specification1}
\widetilde{ \mathbf{\Gamma} }_t 
:= 
\begin{bmatrix}
\gamma_{1,t} &  \widetilde{ \gamma }_{(1,2),t}  & \dots \widetilde{ \gamma }_{(1,N),t} \\
    \vdots & \ddots & \vdots  \\
   \widetilde{ \gamma }_{(N,1),t} &  \dots & \gamma_{N,t} \\
\end{bmatrix}
\ \ \ \text{with} \ \ \ \widetilde{ \gamma }_{(i,j),t} :=\frac{ \gamma_{(i,j),t}+ \gamma_{(j,i),t}}{2}.
\end{align}
such that $\widetilde{ \mathbf{\Gamma} }_t$ is the corresponding the symmetrized risk matrix. In other words, the symmetric risk matrix $\widetilde{ \mathbf{\Gamma} }_t$ given in \eqref{matrix.new.specification1} indicates a matrix structure in which the diagonal elements correspond to the VaR$^{+}$ measure of each asset while the off-diagonal entries represent the weighted pairwise $\Delta \text{CoVaR}$ measures across the set of assets standardized with the term $\left( \gamma_{i,t} \gamma_{j,t} \right)^{1/2}$ for all $i,j \in \left\{ 1,...,  N \right\}$ as defined in expression \eqref{gammamat}.

The intuition behind the particular construction of the risk matrix is to capture Granger causality at the tail of the distributions among pairs of stocks. More specifically, the tail dependence is modelled by the risk measures of VaR and CoVaR which are estimated as tail forecasts based on quantile regression specifications and therefore represent population parameters generated by a cumulative distribution function $F(.)$. The elements on the off-diagonal of the matrix, i.e., $\widetilde{ \gamma }_{(i,j),t} $ are defined to be as  $\widetilde{ \gamma }_{(i,j),t} :=\frac{ \gamma_{(i,j),t}+ \gamma_{(j,i),t}}{2}$, for $i \neq j$, which provide a way to symmetrize the original risk matrix, $\mathbf{\Gamma}_t$. Therefore, these elements represent averaged spillover effects for pair of nodes in the graph. Furthermore, it is not difficult to realize that we are actually proposing a risk matrix with dependence structure that captures \textit{simultaneous causality} (e.g., see \cite{huemer2003causation}) due to the regression based estimation procedure we propose to obtain the elements of the $\mathbf{\Gamma}_t$ matrix. Next, we focus on the utilization of the proposed risk matric in portfolio choice problems. 

\begin{proposition}
\label{propositon1}
The minimization problem given by \eqref{minVaR} is equivalent to minimizing $\mathbf{w}^{\prime} \widetilde{ \mathbf{\Gamma} }_t \mathbf{w}$ subject to the full investment constraint $\mathbf{w}^{\prime} \mathbf{1}= 1$.
\end{proposition}

\newpage 

\begin{proof}
The proof of this result follows from noting that $\mathbf{w}^{\prime} \mathbf{\Gamma}_t \mathbf{w} = \mathbf{w}^{\prime} \widetilde{ \mathbf{\Gamma} }_t \mathbf{w}$. The particular result corresponding to the covariance matrix is proved by \cite{csaki1970concise}.
\end{proof}
Furthermore, the symmetry\footnote{\cite{chen2021asymmetry} study the eigendecomposition for nonsymmetric square matrices. Since we consider the symmetrized version of the risk matrix, thus the estimation procedure for the eigendecomposition is simplified. }  of $\widetilde{ \mathbf{\Gamma} }_t$ entails the spectral decomposition $\widetilde{ \mathbf{\Gamma} }_t = \mathbf{U}_t \mathbf{D}_{\gamma, t} \mathbf{U}_t^{\prime}$, with $\mathbf{U}_t$ an $N \times N$ orthonormal matrix such that $\mathbf{U}_t^{\prime} = \mathbf{U}_t^{-1}$ which includes as columns the linearly independent eigenvectors of $\widetilde{ \mathbf{\Gamma} }_t$, and $\mathbf{D}_{\gamma,t}$ an $N \times N$ diagonal matrix with the corresponding real eigenvalues, defined as $\mathbf{D}_{\gamma} = diag \left\{ \lambda^{ \tilde{\gamma} }_1,..., \lambda^{ \tilde{\gamma} }_N \right\}$. To simplify the notation, we omit the time subscript such that the diagonal matrix and rearrange the diagonal eigenvalue matrix as $\mathbf{D}_{\gamma} =\mathbf{U}^{\prime} \widetilde{ \mathbf{\Gamma} } \mathbf{U}$. Since, in our framework we assume that $N$ remains fixed while $T \to \infty$, then the sample eigenvalues and eigenvectors are consistent estimators of their population counterparts.  

Then, the eigenvalues of the risk matrix $\widetilde{ \mathbf{\Gamma} }$ can be decomposed as a function of the proposed tail risk measures as following
\begin{align} 
\label{lambda1}
\lambda^{ \tilde{\gamma} }_{k} = \overset{N}{\underset{i=1}{\sum}}  u_{ik}^2 \gamma_{k} + \overset{N}{\underset{i=1}{\sum}}  \overset{N}{\underset{\underset{j \neq i}{j=1}}{\sum}} u_{ik} u_{jk} \widetilde{ \gamma}_{(i,j)},
\end{align}
where $\left\{ u_{ij} \right\}_{i,j = 1,...,N}$ denotes the elements of the eigenvectors of the risk matrix $\widetilde{ \mathbf{\Gamma} } \in \mathbb{R}^{N \times N}$. 

\medskip

Assumption \ref{assumption1} below formalize the regulatory condition so that the positive definitiness property of the risk matrix holds based on expression \eqref{lambda1}. Therefore, under the condition provided by Assumption \ref{assumption1} it holds that $\lambda^{ \tilde{\gamma} }_{k} > 0 \ \forall \ k \in \left\{ 1,...,N \right\}$, 
\begin{assumption} 
\label{assumption1}
The positive definite property of the risk matrix is ensured by the conditions
\begin{itemize}
\item[(i)] 
The eigenvectors $\{ u_{ij} \}_{i,j=1,...,N}$ of the risk matrix $\widetilde{ \mathbf{\Gamma} }$ satisfy that 
\begin{equation}
\overset{N}{\underset{i=1}{\sum}} u_{ik}^2 > - \overset{N}{\underset{i=1}{\sum}}  \overset{N}{\underset{\underset{j \neq i}{j=1}}{\sum}} u_{ik} u_{jk} \frac{ \widetilde{ \gamma}_{(i,j)} }{  \gamma_{k} }, \ \ \ \forall \ \ k \in \left\{ 1,...,N \right\}. 
\end{equation}

\item[(ii)] The eigenvalues of the matrix $\widetilde{ \mathbf{\Gamma} }$ are bounded uniformly above and away from zero such that $0 < \lambda_{\text{min}} ( \widetilde{ \mathbf{\Gamma} } ) \leq \lambda_{\text{max}} ( \widetilde{ \mathbf{\Gamma} } )$. 

\item[(iii)] The spectral radius of the matrix is bounded such that $\rho ( \widetilde{ \mathbf{\Gamma} } ) < 1$. 
\end{itemize}

\end{assumption}
\begin{remark}
The condition (i) and (iii) given by Assumption \ref{assumption1} guarantees that the risk matrix $\widetilde{ \mathbf{\Gamma} }$ is positive definite, which implies that the corresponding quadratic form $\mathbf{w}^{\prime} \widetilde{ \mathbf{\Gamma} } \mathbf{w}$ has certain desirable properties for portfolio optimization problems (See, Proposition \ref{propositon2} and \ref{propositon3}). 
Notice also condition (iii) gives a bound for the spectral radius of the matrix that corresponds to the largest eigenvalue of the matrix, which is another restriction that ensures positive definiteness.

\newpage

Furthermore, condition (ii) ensures that the smallest eigenvalue does not move toward zero, to avoid ill-conditioning\footnote{See for example \cite{yu1991recursive}. Notice that a rule of thumb of the occurrence of an ill-conditioned inverse is to have a large value of the ratio $\kappa = \lambda_{\text{max}} \big/ \lambda_{\text{min}}$ (see, \cite{anufriev2015connecting}, \cite{chen2021asymmetry}).} (i.e., the existence of a singular concentration matrix).  
\end{remark}

\begin{remark}
Notice that by definition the risk matrix $( \widetilde{ \Gamma}_{ij} )$ has elements with a certain form of dependence due to the estimation procedure we follow to obtain estimates for the risk measures of VaR and CoVaR as well as the standardization we apply to ensure the positive definiteness of the risk matrix. Therefore, the elements exhibit heavy tails and tail dependence.  
\end{remark}
Next, we focus on the properties of the portfolio risk since within our framework we study an investor who is concerned with tail events, as captured by the tail risk measures of the $\widetilde{ \mathbf{\Gamma} }$ risk matrix. Therefore, the portfolio risk denoted with $\mathcal{Q}( \widetilde{ \mathbf{\Gamma} } ,\mathbf{w})$ shall satisfy the conditions given by Proposition \ref{propositon2} below. 

\begin{proposition}
\label{propositon2}
Let $\mathcal{Q}( \widetilde{ \mathbf{\Gamma} } ,\mathbf{w}) = \mathbf{w}^{\prime} \widetilde{ \mathbf{\Gamma} } \mathbf{w}$ denote the portfolio risk, where the risk matrix $\widetilde{ \mathbf{\Gamma} }$ satisfies Assumption \ref{assumption1}. Then, the following conditions hold
\begin{itemize}
\item[(i)] $\mathcal{Q}( \widetilde{ \mathbf{\Gamma} },\mathbf{w})= 0$ if and only if $\mathbf{w}=0$.
\item[(ii)] $\mathcal{Q}( \widetilde{ \mathbf{\Gamma} },\mathbf{w}) > 0$ for $\mathbf{w} \neq 0$.
\item[(iii)] The associated quadratic form defined by the Lagrangian of the objective function \eqref{minVaR} given by $\mathcal{Q}( \widetilde{ \mathbf{\Gamma} },\mathbf{w}) + \zeta (\mathbf{w}^{\prime} \mathbf{1} - 1)$, for some $\zeta > 0$,  has a global minimum on $\mathbf{w}$.
\end{itemize}
\end{proposition}

\begin{proof}
The proof of these properties follows from standard results on quadratic forms. 
\end{proof}

\medskip

Importantly, under the above conditions, the risk matrix $\widetilde{ \mathbf{\Gamma} }$ is invertible and ensures that a closed-form solution to the portfolio optimization problem given by expression \eqref{minVaR} exists.

\begin{proposition}
\label{propositon3}
The solution to the optimal portfolio allocation problem \eqref{minVaR} with associated matrix $\widetilde{ \mathbf{\Gamma} }$ satisfying Assumption \ref{assumption1} is 
\begin{equation}
\mathbf{w}^{*} = \frac{ \widetilde{ \mathbf{\Gamma} }^{-1} \mathbf{1}}{\mathbf{1}^{\prime} \widetilde{ \mathbf{\Gamma} }^{-1} \mathbf{1} }.   \label{eq:w1}
\end{equation}
\end{proposition}

\begin{proof}
The proof of this proposition is immediate from the first order conditions of the maximization problem \eqref{minVaR}.  
\end{proof}

\bigskip

Notice that Proposition \ref{propositon3} ensures that using our proposed risk matrix in optimal portfolio choice problems a corresponding closed form solution to the traditional Markowitz portfolio optimization problem exists. Furthermore, in this paper we are interested to obtain some further insights regarding the network topology and how the tail connectedness of nodes affects the optimal portfolio allocation as well as the portfolio risk function of the investor. 

\newpage 

In summary, in this section we introduce a novel risk matrix which is constructed based on tail risk measures. We showed that the proposed risk matrix has a suitable specification for optimal portfolio allocation problems. In Section \ref{SectionII}, we assume that the stocks in the portfolio represent nodes in a network and proposed a more formal framework which can be utilized as a mechanism for portfolio selection based on graphs under tail events.    

Another important aspect is the consistent estimation of the risk matrix $\widetilde{ \mathbf{\Gamma} }$. Even though we do not formally examine this in this paper, and leave this for a future study, our proposed risk matrix is suitable for high dimensional settings as well. However, the computation time for estimation of the inverse of the risk matrix, that is, $\widetilde{ \mathbf{\Gamma} }^{-1}$, can be affected by the number of stocks in the portfolio. In that case, we recommend the use of a pseudoinverse estimation procedure such as the implementation of the Moore–Penrose matrix. Other more sophisticated estimation procedures include for example the constrained $\ell_1-$minimization for inverse matrix estimation (CLIME) proposed in  \cite{cai2011constrained} (see also \cite{cai2020high}). Furthermore, the dependence structure of our risk matrix is a further aspect worth examining in future research  (e.g., see the framework proposed by \cite{shu2019estimationmatrix}). 

Notice that in the literature of large covariance matrices as well as in graph based modelling approaches, (e.g., as in \cite{anufriev2015connecting}), the partial correlation between      random variables $X_i$ and $X_j$ given $\left\{ X_k, k \neq i, j \right\}$ is equal to $- \omega_{ij} / \sqrt{\omega_{ii} \omega_{jj}}$. Therefore, zero partial correlation means conditional independence between Gaussian random variables. Within our framework, zero tail dependence or tail connectivity due to the underline graph structure we impose, implies that there is a common level of Value-at-Risk which affects the nodes. Next, we focus on the relationship between the vector of optimal weights, that is, the optimal portfolio allocation given by expression \eqref{eq:w1}, and the dependence in the tails between the stocks of the portfolio. In particular, we utilize the notion of eigenvector centrality and by applying the spectral vector decomposition we derive an analytical expression for the eigenvalues of the risk matrix with respect to the risk measures of VaR and CoVaR. 

\begin{example}
\label{Example1}
Consider a portfolio with $N = 3$ stocks. An investor who has tail concerns he invests in the portfolio based on the tail dependence across the underline distribution of returns represented as nodes in the graph, as captured by the VaR-CoVaR risk matrix defined  
\begin{align}
\widetilde{ \mathbf{\Gamma} } = 
\begin{bmatrix}
\gamma_{1} &  \widetilde{ \gamma }_{(1,2)}  & \widetilde{ \gamma }_{(1,3)} 
\\
\widetilde{ \gamma }_{(2,1)} &  \gamma_{2}  & \widetilde{ \gamma }_{(2,3)} 
\\
\widetilde{ \gamma }_{(3,1)} &  \widetilde{ \gamma }_{(3,2)}  &\gamma_{3} 
\end{bmatrix}
\end{align} 
Therefore, the optimal decision problem entails to find the optimal portfolio weight based on the full investment constraint and a positive weight vector constraint. Denote with $\widetilde{k}_{ij}$ the elements of the precision matrix of VaR-$\Delta$CoVaR, where $\widetilde{ \mathbf{K} } \in \mathbb{R}^{N \times N}$ such that $\widetilde{ \mathbf{K} } = \widetilde{ \mathbf{\Gamma} }^{-1}$. This implies that $\widetilde{ \mathbf{\Gamma} }^{-1} . \mathbf{1} \equiv \widetilde{ \mathbf{K} } . \mathbf{1}$ represents a vector with the same dimensions as the optimal portfolio weight, which has elements the row-sums of the precision matrix $\widetilde{ \mathbf{K} }$. 

\newpage 

In vector form is expressed as below  
\begin{align}
\widetilde{ \mathbf{\Gamma} }^{-1} . \mathbf{1} = \left( \sum_{j=1}^N \widetilde{k}_{1j}, \sum_{j=1}^N \widetilde{k}_{2j} , \sum_{j=1}^N  \widetilde{k}_{3j} \right)^{\prime}
\end{align}  
Moreover, the denominator of the optimal portfolio vector is  $\mathbf{1}^{\prime} . \widetilde{ \mathbf{K} } . \mathbf{1}$ which gives the sum of the elements of the precision matrix. We find that the optimal weight vector is expressed
\begin{align}
\mathbf{w}^{*} = \frac{1}{ \displaystyle \sum_{j=1}^N \widetilde{k}_{1j} + \sum_{j=1}^N \widetilde{k}_{2j} + \sum_{j=1}^N \widetilde{k}_{3j} } \left( \sum_{j=1}^N \widetilde{k}_{1j}, \sum_{j=1}^N \widetilde{k}_{2j} , \sum_{j=1}^N  \widetilde{k}_{3j} \right)^{\prime}
\end{align}
\end{example}
Notice that Example \ref{Example1} above provides some useful insights regarding the key ideas of our framework. The dependence of the optimal portfolio weights to the tail risk measures of VaR and CoVaR indicate that diversification occurs by balancing-out the tail dependence across the nodes of the graph. In particular, by diversifying based on tail connectivity allows us to consider the optimal portfolio allocation problem\footnote{Notice that in contrast to other framework such as the naive equal-weighted scheme examined by \cite{nguyen2020investigating}, in the context of tail-risk dependence, we consider the minimum variance portfolio optimization which permits to study the relation optimal asset allocation and stock centrality.} under further constraints such as the target expected returns as the mean-variance portfolio. Taking into consideration the co-movements of the market during extreme events by incorporating the tail dependence structure across the graph extends the traditional variance-covariance diversification strategy. 

Next, we impose an additional assumption to control the sparsity of the precision matrix (see, \cite{callot2021nodewise}).  Denote with $\mathcal{S}_i := \left\{ j \ \text{such that} \ \widetilde{k}_{ij} \neq 0 \right\}$ to be the set of nonzero parameter estimates of the precision matrix for the $i-$th row, $\widetilde{k}_{i}$, and denote with $s_i := | \mathcal{S}_i |$ to be its cardinality (the number of elements of the particular set). Let $\bar{s} := \underset{ 1 \leq i \leq p }{ \text{max} } s_i$, where $s_i$ represents the number of nonzero parameters in the $i-$th row. 

\begin{assumption}
\label{sparsity.assumption}
The following condition holds 
\begin{align}
\bar{s} \sqrt{ \text{log} \left( N / T \right) } = o(1). 
\end{align}
\end{assumption}

Assumption \ref{sparsity.assumption} provides a sparsity condition for the inverse of the risk matrix of VaR-$\Delta$CoVaR. However, notice that the sparsity of the precision matrix does not necessarily imply sparsity of the VaR-$\Delta$CoVaR matrix. In particular, the weakly dependence structure with possibly temporally or correlated parameter estimates produce a risk matrix which is not necessarily sparse even though the entries of the risk matrix from a data-based estimation can include near-zero terms, especially for certain nodes in the graph. For simulation-based parameter estimation the entries of the matrix is likely to have a larger $\bar{s}$ value.

\newpage 

\section{Portfolio Choice under Tail Dependence in Graphs} 
\label{SectionII}

In this Section we examine the mechanism of optimal portfolio choice under the assumption of network tail dependence. Related studies to tail dependence measures and assumptions are \cite{escanciano2019measuring}, \cite{zhang2021high} while   \cite{callot2019nodewise} consider the portfolio choice problem. Firstly, we examine the relationship between tail connectivity and stock centrality, focusing on two aspects: (i) relation between stock centrality and  portfolio risk; and (ii) relation between stock centrality and optimal portfolio allocation. Secondly, we show how the procedure can be applied to affine transformation of the proposed tail risk matrix. 

\subsection{Tail Connectivity and Stock Centrality}
\label{SectionII.A}

We denote with $ \mathcal{G} = \{ \mathcal{V} , \mathcal{E} \}$ a graph structure, representing a network that consists of a set of nodes, $\mathcal{V} \in \left\{ 1,...,N \right\}$, and a set of edges, $\mathcal{E}$, connecting the pairs of nodes. A full characterization of the information in the network is provided by the $N \times N$ adjacency matrix $\mathbf{\Omega}$ whose element $\{ \Omega_{ij} \}_{i,j=1}^{N}$ determine the existence of a link connecting node $i$ and $j$ for the graph $\mathcal{G}$. The link between two elements can be characterized by a binary variable which determines the existence of a connection or by a real value that corresponds to the pairwise weight between nodes. We propose the following adjacency matrix $\mathbf{\Omega} = \widetilde{ \mathbf{\Gamma} } - diag ( \widetilde{ \mathbf{\Gamma} } )$, where $diag(\cdot)$ is an $N \times N$ diagonal matrix, implying that $diag ( \widetilde{ \mathbf{\Gamma} } ) :=  \text{diag} \{ \gamma_1, \ldots, \gamma_N \}$. 

Therefore, within our framework we consider a weighted closed network, where the links between stocks are determined by the $\widetilde{ \gamma}_{(i,j)}$ measures as defined by the off-diagonal elements of the $\widetilde{ \mathbf{\Gamma} }$ matrix. The adjacency matrix $\mathbf{\Omega}$ is symmetric by definition, since the main diagonal has a vector of zeros and the off-diagonal terms are the elements of the symmetric risk matrix $\widetilde{ \mathbf{\Gamma} }$, that is, $\Omega_{ij} \equiv \widetilde{ \gamma}_{(i,j)}$ for $i \neq j$. Furthermore, the symmetry of the adjacency matrix entails the spectral decomposition $\mathbf{\Omega} = \mathbf{Z}_{ \Omega } \mathbf{D}_{ \Omega} \mathbf{Z}_{ \Omega }^{\prime}$, with $\mathbf{Z}_{ \Omega }$ an $N \times N$ orthonormal matrix such that $\mathbf{Z}_{ \Omega }^{\prime} = \mathbf{Z}_{ \Omega }^{-1}$ that contains the linearly independent eigenvectors of $\mathbf{\Omega}$, and $\mathbf{D}_{\Omega}$ is an $N \times N$ diagonal matrix with the corresponding eigenvalues. Thus, expressing the eigenvalues of the adjacency matrix $\mathbf{\Omega}$ using the spectral vector decomposition we obtain   
\begin{align} 
\label{lambda1b}
\lambda_{k}^{ \Omega } =\sum_{i=1}^{N} \overset{N}{\underset{\underset{j \neq i}{j=1}}{\sum}} z_{ik} z_{jk} \widetilde{ \gamma}_{(i,j)},
\end{align}
with $z_{ij}$ the eigenvectors of the matrix $\mathbf{\Omega} \in \mathbb{R}^{N \times N}$. 

\medskip

We decompose the loss function $\mathcal{Q}( \widetilde{ \mathbf{\Gamma} } ,\mathbf{w}) = \mathbf{w}' \widetilde{ \mathbf{\Gamma} } \mathbf{w}$  as the sum of a quadratic loss function given by the adjacency matrix and a quadratic loss function given by a diagonal risk matrix with elements given by $\gamma_i$. More formally,
an equivalent expression for the quadratic form is 
\begin{equation} 
\label{decom}
\mathcal{Q}( \widetilde{ \mathbf{\Gamma} } ,\mathbf{w}) := \mathbf{w}^{\prime} \widetilde{ \mathbf{\Gamma} }
 \mathbf{w} = \left\{ \mathbf{w}^{\prime} \mathbf{\Omega} \mathbf{w} + \sum_{i=1}^{N} w_{i}^2 \gamma_i \right\}.
\end{equation}

\newpage 

Proposition \ref{proposition4} below provides a theoretical result regarding the relation between the asset allocation and the stock centrality within the financial network.

\medskip

\begin{proposition}  
\label{proposition4}
For a given vector $\mathbf{w}$, a sufficient condition to guarantee that the portfolio risk $\mathcal{Q}( \mathbf{ \Omega },\mathbf{w}) = \mathbf{w}^{\prime} \mathbf{ \Omega } \mathbf{w}$ is increasing on the magnitude of the tail risk measure $\gamma_{i|j}$ is that the derivative of $\lambda_{k}^{ \Omega }$ with respect to $\widetilde{\gamma}_{i|j}$ is greater than zero for all $i,j,k \in \left\{ 1, \ldots, N \right\}$.
\end{proposition}

\bigskip

We begin our analysis, by studying the contribution of stock centrality to the portfolio risk given by $\mathcal{Q}( \widetilde{ \mathbf{\Gamma}} , \mathbf{w})$. Therefore, by considering expression \eqref{decom} as well as the decomposition of the quadratic form $\mathcal{Q}( \mathbf{ \Omega },\mathbf{w}) = \mathbf{w}^{\prime} \mathbf{ \Omega } \mathbf{w}$ we can express the eigenvector centrality measure with respect to the spectral vector decomposition of the adjacency matrix $\mathbf{ \Omega }$. A centrality measure quantifies the influence of certain nodes in a given graph. Several centrality measures have been proposed in the literature which consider different features of the network topology. In this paper, we focus on the measure of eigenvector centrality which is closely related to the definition of Katz centrality (see, \cite{katz1953new} and \cite{Bonacich72}). In particular, the eigenvector centrality of stock $i$ is proportional to the weighted sum of its neighbors' centralities. Therefore, it captures the idea that central nodes are those with stronger connections to other central nodes (see, the definition given by \cite{elliott2019network}).  

More formally, for the adjacency matrix $\mathbf{ \Omega }$, the eigenvector $v_i$ is given by the expression
\begin{align} 
\label{central0}
v_i = \left( \lambda^{ \Omega }_{(1)} \right)^{-1} \overset{N}{\underset{j=1}{\sum}} \Omega_{ij} \hspace{0.2ex} v_j,
\end{align}
is a centrality measure associated to node $i$ by the Perron-Frobenius theorem (Appendix \ref{AppendixD}), where $\lambda^{ \Omega }_{(1)}$ is the largest eigenvalue of $\mathbf{\Omega}$ (or the spectrum radius of the adjacency matrix). 

\medskip

Next we investigate the relation between stock centrality and portfolio risk as this aspect can be useful in portfolio choice problems for tail risk events. Theorem \ref{theorem1} shows that, in general, no monotonic relationship between stock centrality and portfolio risk exists.

\begin{theorem}  
\label{theorem1}
The portfolio risk given by the quadratic form  $\mathcal{Q}(\widetilde{\mathbf{\Gamma}},\mathbf{w})$ is not monotonically increasing or decreasing with respect to the centrality measure $v_i$.
\end{theorem}

\medskip

The result given by Theorem \ref{theorem1} indicates that the quadratic form which characterize the optimal investment decision is not monotonic on the centrality of an asset. In other words, investors who optimize using the proposed risk matrix constructed with tail risk measures based on exogenous variables affecting the network topology in the graph, cannot for certain know a priori whether the network topology of assets can directly affect the performance of the portfolio risk, especially when there are no further restrictions regarding short and long positions as reflected by positive and negative signs of the weight allocation.  

\newpage 

For instance, as shown by the proof of Theorem \ref{theorem1} (see Appendix \ref{AppendixA}), the opposite result would follow, that is, we can conclude that a monotonic relation for the portfolio risk holds with respect to the corresponding centrality measure $v_i$ if and only if we allow the investment portfolio to take negative positions on some stocks, \textit{i.e.} $w_i <0$ for some $i \in \left\{ 1, \ldots, N \right\}$, since the maximum eigenvalue of the adjacency matrix $\mathbf{\Omega}$ cannot be negative. Therefore, with the additional investment constraint of long positions and by ensuring that the adjacency matrix $\mathbf{\Omega}$ to be positive definite such that the eigenvector centrality satisfies $\mathbf{v}=(v_1, \ldots, v_N) \geq 0$, then the result given by Theorem \ref{theorem1} can be refined. We introduce more formally these conditions with the following assumption.

\medskip

\begin{assumption} 
\label{assumption2}
Let $\mathbf{\Omega}$ be positive definite and such that the eigenvector centrality given by \eqref{central} satisfies $\mathbf{v} \geq 0$. 
\end{assumption}

Next, Corollary \ref{corollary1} below gives the condition for a monotonic relation between portfolio risk function which is almost surely positive-definite and the stock centrality measure corresponding to the symmetrized risk matrix. Notice that Corollary \ref{corollary1} only holds under the additional condition we impose by Assumption \ref{assumption2}.

\begin{corollary} 
\label{corollary1}
Under Assumption \ref{assumption2}, the portfolio risk defined by $\mathcal{Q}(\widetilde{\mathbf{\Gamma}},\mathbf{w})$ is increasing on the centrality measure $v_i$ if the portfolio weights are all non-negative, \textit{i.e.} $\mathbf{w} \geq 0$.
\end{corollary}

\begin{proof}
The proof of this result is immediate from expression \eqref{derlambda} given in Appendix \ref{AppendixA}.
\end{proof}

\bigskip

\begin{remark}
Notice that  Corollary \ref{corollary1} implies that a simple heuristic for having a portfolio risk that is kept in low levels is to have a centrality vector with lower values in comparison to a portfolio with high centrality, which implies a high level of tail connectedness among the nodes in the network. Moreover, the particular result holds under the assumption that the investor chooses to invest in all stocks in the portfolio.  
\end{remark}

Therefore, when the investor with tail risk concerns decides to exclude central stocks from the portfolio with graph representation sequentially with a short-selling investment constraint only, then the portfolio risk function can be maintained in lower levels than investing to all stocks without considering separately the effect of  the network topology. Moreover, in this paper we avoid to examine further the relationship between optimal portfolio allocation and stock centrality as the literature currently has mixed results on the particular aspect. More specifically, \cite{peralta2016network} concludes that optimal portfolio strategies should underweight the allocation to high central assets and overweight the allocation to low central assets. A similar finding is demonstrated by  \cite{huttner2016portfolioselection}. On the other hand, the study of \cite{Olmo2021AssetCentr} concludes that higher asset centrality implies a larger allocation on the risky assets. Notice that both these results correspond to the case of the correlation matrix as an adjacency matrix, therefore a further investigation is necessary to examine whether similar findings can be established for our proposed risk matrix, which captures tail dependence. 

\newpage 

\subsubsection{Optimal Asset Allocation and Stock Centrality}

In this subsection, we examine the relation between optimal asset allocation and stock centrality for the optimal portfolio choice problem under tail risk events. To do so, we express the risk matrix as below
\begin{align}
\widetilde{ \mathbf{ \Gamma} } \equiv \big( \mathbf{I}_N -  \widetilde{ \mathbf{\Omega} } \big), \ \text{where} \ \ \widetilde{\mathbf{\Omega}} = - \left( \mathbf{ B } + \mathbf{\Omega} \right),
\end{align}
such that $\mathbf{B} := diag ( \widetilde{\mathbf{\Gamma}} ) - \mathbf{I}_N$. 

The above trick allow us to express the vector of optimal portfolio weight vector $\mathbf{w}^*$ as a function of the eigenvector centrality measure. Notice that, the matrix $\widetilde{ \mathbf{\Omega}}$ inherits the properties of the adjacency matrix $\mathbf{\Omega}$, hence, it is symmetric. The diagonal elements of $\mathbf{\Omega}$ are equal to $\left( 1- \gamma_k \right)$ for all $k \in \left\{ 1, \ldots , N \right\}$. 

Then, by applying the spectral vector decomposition, we obtain $\widetilde{ \mathbf{ \Omega}} = \mathbf{S} \mathbf{D}_{ \widetilde{ \Omega} } \mathbf{S}^{\prime}$,
with $\mathbf{S}$ an orthonormal matrix comprised by the eigenvectors of the matrix $\widetilde{ \mathbf{ \Omega}}$, and $\mathbf{D}_{ \widetilde{ \Omega} }$ a diagonal matrix with elements given by the corresponding eigenvalues $\lambda_k^{ \widetilde{\Omega} }$. Thus, the centrality measure that corresponds to the matrix $\widetilde{ \mathbf{ \Omega }}$ is given in Corollary \ref{corollary21} below.

\begin{corollary}
\label{corollary21}
The eigenvector centrality measure $\widetilde{v}_i$ associated to the matrix $\widetilde{ \mathbf{ \Omega } }$ is defined as 
\begin{align} 
\label{central1}
\widetilde{v}_i = \left( \lambda^{ \widetilde{\Omega} }_{(1)} \right)^{-1}  \overset{N}{\underset{j=1}{\sum}} \widetilde{\Omega}_{ij} \hspace{0.2ex} \widetilde{{v}}_j ,
\end{align}
where $\lambda^{ \widetilde{ \Omega} }_{(1)}$ is the largest eigenvalue of the adjacency matrix $\widetilde{ \mathbf{\Omega}}$, such that the eigenvalues follow the ordering $\lambda_{ \text{max} }^{ \widetilde{\Omega} } \equiv \lambda_{(1)}^{ \widetilde{\Omega} } \geq \ldots \geq \lambda_{(N)}^{ \widetilde{\Omega} } \equiv  \lambda_{ \text{min} }^{ \widetilde{\Omega} }$. 
\end{corollary}

We are now ready to introduce the main result of this section, namely, that under general conditions higher asset centrality implies a larger allocation on the risky asset. This result contrasts with recent insights in the related literature, see \cite{peralta2016network}, showing that optimal strategies should underweigh the allocation to high central assets and overweigh the allocation to low central assets within a optimal portfolio choice problem under tail events. Intuitively speaking, we find analytical evidence that for investors who are concern with tail risk and the level of connectivity in the network as it is captured by tail dependence across the nodes, it is optimal to invest more in higher centrality stocks rather than low centrality stocks. The particular heuristic obviously refer to the minimum-variance portfolio optimization where the variance-covariance matrix is replaced with the VaR-$\Delta$CoVaR matrix, and therefore it does not necessarily reflect what is the effect of stock centrality within the given network topology to portfolio risk captured again through the tail risk measures we consider in this paper. Specifically, we examine the latter aspect via our empirical implementation and in the particular our proposed algorithm for determining the optimal number of stocks based on the tail centrality measure. Nevertheless, in this subsection we provide the related analytical results for the relation between optimal asset allocation and stock centrality.   

\newpage 

To formally present our analytical results in propositions we additionally impose the conditions given by Assumption \ref{assumption3} and \ref{assumption4}. 

\begin{assumption} 
\label{assumption3}
Let $\{ \lambda_i^{ \widetilde{\Omega} } \}_{i=1,...,N}$ be the eigenvalues of the adjacency matrix $\widetilde{ \mathbf{\Omega}}$. Then, we assume that $0 <\lambda_i^{ \widetilde{\Omega} } < 1$ for all $ i \in \left\{ 1, \ldots, N \right\}$.
\end{assumption}

The condition given by Assumption \ref{assumption3} guarantees that the matrix $\widetilde{ \mathbf{ \Omega }}$ is positive definite and with spectral radius less than one. Under this condition, we obtain the following expression for the optimal portfolio weights.

\medskip

\begin{proposition}
\label{proposition7}
Under Assumption \ref{assumption1} - \ref{assumption3}, the optimal portfolio allocation $w_i^*$ in \eqref{eq:w1} can be expressed as a function of stock centrality as below
\begin{align} 
\label{eq:w1a}
w_i^{*} 
= \frac{ \displaystyle \frac{ \widetilde{{v}}_{i} }{1- \lambda_{(1)}^{ \widetilde{\Omega} }  } \left(\overset{N}{\underset{j=1}{\sum}} \widetilde{ {v}}_{j} \right) + \overset{N}{ \underset{k=2}{\sum}} \frac{s_{ik}}{1- \lambda_k^{ \widetilde{ \Omega} } } \left(\overset{N}{\underset{j=1}{\sum} } s_{jk} \right)}{  \displaystyle \frac{1}{1- \lambda_{(1)}^{ \widetilde{ \Omega} }  } \left(\overset{N}{\underset{j=1}{\sum}} \widetilde{v}_{j} \right)^2 + \overset{N}{\underset{k=2}{\sum}} \frac{1}{1- \lambda_k^{ \widetilde{\Omega} }   } \left(\overset{N}{\underset{j=1}{\sum}} s_{jk} \right)^2},
\end{align}

with $\lambda_{(1)}^{ \widetilde{\Omega} }$ the largest eigenvalue of $\mathbf{ \widetilde{ \Omega} }$ and $\widetilde{ \mathbf{v}}$ the corresponding eigenvector centrality vector of the particular adjacency matrix.
\end{proposition}

\medskip

The result in Proposition \ref{proposition7} allows us to explore further the relationship between an asset's centrality position in a portfolio and its optimal portfolio allocation. The following result provides sufficient conditions for the existence of a positive relationship between the optimal asset allocation and its centrality in the portfolio. To show this, we need the following technical conditions that we present as an additional assumption.

\medskip

\begin{assumption} 
\label{assumption4}
Let $\left\{ s_{jk} \right\}_{j,k=1,...,N}$ be the elements of the eigenvectors of the adjacency matrix $\widetilde{ \mathbf{\Omega}}$, and let $\left\{ \widetilde{v}_{j} \right\}_{j=1,...,N}$ be the corresponding eigenvector centrality measures. Then, we impose the following conditions:
\begin{itemize}
\item[i)] $\overset{N}{\underset{j=1}{\sum}} s_{jk}  > 2 s_{ik}$, for $i,k=1,\ldots,N$; 
\item[ii)] $\overset{N}{\underset{j=1}{\sum}} \widetilde{v}_{j} > 1$;
\item[iii)] $\widetilde{v}_{i} \left(\overset{N}{\underset{j=1}{\sum}} s_{jk} \right) > \displaystyle \frac{ s_{ik} }{ 1 - \lambda_{(1)}^{ \widetilde{\Omega} } }$, for $i,k \in  \left\{ 1,\ldots,N \right\}$.
\end{itemize}
\end{assumption}

The conditions given by Assumption \ref{assumption4} are sufficient but not necessary conditions, and can be relaxed at the expense of more algebra as explained in the proof of the result in the mathematical appendix.

\newpage 

\begin{theorem}
\label{theorem2}
Under Assumptions \ref{assumption3}-\ref{assumption4}, the optimal portfolio allocation $w_i^*$ in \eqref{eq:w1a} increases as the corresponding eigenvector centrality $\widetilde{ {v} }_i$ increases.
\end{theorem}

\medskip

The above result can be illustrated in more detail if we assume that the VaR measures are common across assets, that is, $\gamma_i = \gamma$, for all $i \in \left\{ 1, \ldots, N \right\}$. In this scenario, we can refine the result in Theorem \ref{theorem1} and show the existence of a positive relationship between the allocation to the risky asset and the centrality measure $v_i$ defined in \eqref{central}. More formally, this implication of Theorem \ref{theorem2} is presented by Proposition \ref{proposition8} below, which holds in the case we assume that same Value-at-Risk corresponds to all assets in the financial network. 

\medskip

\begin{proposition}
\label{proposition8}
Under Assumptions \ref{assumption3}-\ref{assumption4} and a common $\gamma$ across risky assets, the optimal portfolio vector $w_i^*$ in \eqref{eq:w1} increases as the corresponding eigenvector centrality $v_i$ increases.
\end{proposition}

The proof of Proposition \ref{proposition8} can be found in  Appendix \ref{AppendixA} of the paper. Moreover, using previous arguments, it is straightforward to see that when $\gamma_i = \gamma$ for all $i \in \left\{ 1, \ldots, N \right\}$ then $\lambda_i^{ \widetilde{ \Omega} } = \left( 1 - \gamma - \lambda_i^{ \Omega } \right)$ such that the optimal portfolio allocation becomes
\begin{equation} 
\label{weight6}
\mathbf{w}_i^{*} 
= \frac{ \displaystyle \frac{v_{i} }{ \gamma + \lambda_{(1)}^{ \Omega } } \left( \overset{N}{\underset{j=1}{\sum}} v_{j} \right) + \overset{N}{ \underset{k=2}{\sum}} \frac{s_{ik}}{ \gamma + \lambda_k^{ \Omega }} \left(\overset{N}{ \underset{j=1}{\sum} } s_{jk} \right) }{ \displaystyle \frac{1}{ \gamma + \lambda_{(1)}^{ \Omega } } \left( \overset{N}{\underset{j=1}{\sum}} v_{j} \right)^2 + \overset{N}{\underset{k=2}{\sum}} \frac{1}{ \gamma + \lambda_k^{ \Omega } } \left(\overset{N}{ \underset{j=1}{\sum} } s_{jk} \right)^2},
\end{equation}
with $\lambda_{i}^{ \Omega}$ the eigenvalues of the adjacency matrix $\mathbf{ \Omega }$. 

\medskip

In particular expression \eqref{weight6} provides a suitable parametrization of the optimal weight vector for proving the relation between optimal portfolio allocation and stock centrality. Similarly to the above analysis one can obtain the corresponding optimal allocation vector for the mean VaR-$\Delta \text{CoVaR}$ portfolio\footnote{To construct the particular portfolio optimization problem, one needs to consider the objective function along with the target expected mean constraint and show that there is an equivalent closed-form solution to the case of the covariance matrix. A lot of applications of this framework can be found in the literature.} problem. We omit the particular derivations from this section, however the dependence of the optimal weight vector to both the eigenvector centrality and the common VaR measure across assets, still holds. In summary, the main result and novel contribution to the literature as demonstrated with our analysis in Section \ref{SectionII} is that within the optimal portfolio choice problem for an investor who is concerned with tail risk, higher stock centrality, as measured by eigenvector centrality implies larger allocation on the risky asset. More specifically, due to the structure of our novel risk matrix, the optimal portfolio allocation can be interpreted with respect to the tail risk measures while stock centrality can be interpreted as a tail risk eigenvector centrality, capturing this way the induced centrality of stocks due to comovements in higher-order moments of the underline distributions that occur since our framework considers the effect of tail dependence. 

\newpage

\section{Node Exclusion and Portfolio Risk} 
\label{SectionIII}

The analysis in Section \ref{SectionII} provides conditions under which stock centrality increases the overall risk in the portfolio. The particular  measure is directly related to the magnitude of the $\Delta \text{CoVaR}$ measures and, hence, to the presence of network tail dependence across the nodes. In this context, it may be important to assess the effect of stock centrality to the centrality of other stocks in the graph. Furthermore, even though we do not explicitly examine the effect of stock centrality on optimal portfolio allocation, our main concern in this section is to examine the effect of stock centrality on portfolio risk. The scope of Section \ref{SectionIII} is to propose a suitable algorithm which can eliminate the central assets based on a stopping rule. This technique allows for example to measure the effects of interventions in networks (e.g., see \cite{elliott2019network}, \cite{galeotti2020targeting} and \cite{Badev2021}). 

An initial step towards this direction is to compare stock centrality between two portfolios. The first portfolio is constructed from $N$ assets and characterized by the risk matrix $\widetilde{\mathbf{\Gamma}}$. The second construction is given by a reduced portfolio in which the most central asset has been removed and is  characterized by the risk matrix $\widetilde{\mathbf{\Gamma}}_{\backslash \{k\}}$. To construct this matrix we remove the $k-$th row and $k-$th column, where $k$ denotes the node with the largest eigenvector centrality, that is, $v_k$ is the largest element of the centrality vector $\mathbf{v}= \left( v_1, \ldots , v_N \right)$. Using the matrix decomposition given by \eqref{decom} to the reduced risk matrix $\widetilde{\mathbf{\Gamma}}_{\backslash \{k\}}$, we obtain the following expression for the quadratic form of the portfolio risk induced after removing the $k-$th node from the graph
\begin{equation} 
\label{decom2}
\mathcal{Q}(\widetilde{\mathbf{\Gamma}}_{\backslash \{k\}},\mathbf{w}_{\backslash \{k\}}^*) 
= 
\mathbf{w}_{\backslash \{k\}}^{ * \prime } \mathbf{ \Omega }_{\backslash \{k\}} \mathbf{w}_{\backslash \{k\}}^* + \overset{N}{\underset{\underset{i \neq k}{i=1}}{\sum}} \mathbf{w}_{i \backslash \{k\}}^{*2} \gamma_i,
\end{equation}
where $\mathbf{\Omega}_{ \backslash \{k\} }$ is the adjacency matrix associated to the reduced risk matrix $\widetilde{\mathbf{\Gamma}}_{\backslash \{k\}}$ and $\mathbf{w}_{\backslash \{k\}}^{*}$ are the optimal portfolio weights of the reduced portfolio. Therefore, an expression for the difference of the two loss functions defined as $\mathcal{Q}(\widetilde{\mathbf{\Gamma}},\mathbf{w}^*) - \mathcal{Q}(\widetilde{\mathbf{\Gamma}}_{\backslash \{k\}},\mathbf{w}_{\backslash \{k\}}^*)$ is given by
\begin{equation*}
\label{decomdif}
\begin{aligned}
\mathcal{Q}(\widetilde{\mathbf{\Gamma}},\mathbf{w}^*) - \mathcal{Q}(\widetilde{\mathbf{\Gamma}}_{\backslash \{k\}},\mathbf{w}_{\backslash \{k\}}^*) & = \mathbf{w}^{* \prime } \mathbf{\Omega} \mathbf{w}^{*} - \mathbf{w}_{\backslash \{k\}}^{* \prime } \mathbf{ \Omega }_{\backslash \{k\}} \mathbf{w}_{\backslash \{k\}}^{*} + \overset{N}{\underset{\underset{i \neq k}{i=1}}{\sum}} \big( w_{i}^{*2} -w_{i\backslash \{k\}}^{*2} \big) \gamma_i + w_{k}^{*2} \gamma_k.   
\end{aligned}
\end{equation*} 
The above quantity due to the difference of the loss functions occurs due to the effect of node removal from the graph and thus the level of connectivity in the induced network topology. Then, by expression \eqref{lossgamma.appendix} in Appendix \ref{AppendixA} we obtain that the middle term above becomes
\begin{align}
\label{newexpression}
\mathbf{w}_{\backslash \{k\}}^{* \prime} \mathbf{\Omega}_{\backslash \{k\}} \mathbf{w}_{\backslash \{k\}}^{*} 
=
\mathbf{w}_{\backslash \{k\}}^{* \prime } \mathbf{Z}_{\backslash \{k\}} \mathbf{D}_{ \Omega_{\backslash \{ k \} } } \mathbf{Z}_{\backslash \{k\}}^{ \prime } \mathbf{w}_{\backslash \{k\}}^{*}  
= 
\overset{N}{ \underset{\underset{i \neq k}{i=1}}{\sum}} \lambda_{i\backslash \{k\}}\left(\overset{N}{\underset{j=1}{\sum}} w_{j_{\backslash \{k\}}}^* z_{ji_{\backslash \{k\}}}  \right)^2
\end{align}

\newpage 

Then, by substituting expression \eqref{newexpression} we obtain that
\begin{equation}
\begin{aligned}  
\label{l1}
\mathbf{w}^{* \prime } \mathbf{ \Omega } \mathbf{w}^{*} - \mathbf{w}_{\backslash \{k\}}^{* \prime } \mathbf{ \Omega }_{ \backslash \{k\}} \mathbf{w}_{\backslash \{k\}}^{*} 
       & = \overset{N}{\underset{\underset{i \neq k}{i=1}}{\sum}} \big( \lambda_{i}^{\Omega } - \lambda_{i \backslash \{k\}}^{\Omega } \big) \left(\overset{N}{\underset{\underset{j \neq k}{j=1}}{\sum}}  w_j^* z_{ji}  \right)^2   \\ 
       & + \overset{N}{\underset{\underset{i \neq k}{i=1}}{\sum}} \lambda_{i \backslash \{k\}}^{\Omega }  \left\{ \left(\overset{N}{\underset{\underset{j \neq k}{j=1}}{\sum}} w_j^* z_{ji}  \right)^2 - \left(\overset{N}{\underset{\underset{j \neq k}{j=1}}{\sum}} w_{j \backslash \{k\}}^* z_{ji\backslash \{k\}}  \right)^2 \right\} \\ 
       &   + \overset{N}{\underset{\underset{i \neq k}{i=1}}{\sum}} \lambda_{i}^{\Omega }  \bigg\{ \left( w_k^* z_{ki} \right)^2 + 2 \overset{N}{\underset{\underset{j \neq k}{j=1}}{\sum}} w_k^* w_j^* z_{ki}z_{ji} \bigg\} \\ 
      & + \lambda_{k}^{\Omega } \left( \overset{N}{\underset{j=1}{\sum}} w_j^* z_{jk} \right)^2.
\end{aligned}
\end{equation}

The first and second terms of expression \eqref{l1} reflect differences in eigenvalues and eigenvectors between the two adjacency matrices. That is, the same stock can take different values of the centrality statistic across the two portfolios. However, as \cite{van2014graph} emphasizes the removal of a node $k$ from the graph has a dominant effect on the $\lambda_{\backslash \{k\}}^{\Omega }$ rather on the eigenvectors of the sub-graph. Nevertheless, this effect is weighted by the gap between the eigenvalue of the full matrix, $\lambda^{\Omega }$ and the eigenvalue of the reduced matrix, $\lambda_{\backslash \{k\}}^{\Omega }$. The last two terms capture the contribution of the most central node $k$ to the full adjacency matrix. Moreover, expression \eqref{l1} allows us to disentangle the effect of stock $k$ from the remaining stocks based on the difference of these two quadratic forms. This is done by removing the last two terms in the above expression. Thus, the net difference after discounting the effect of stock $k$, where $k \in \left\{ 1,..., N \right\}$ is denoted by $\Delta$ and defined as
\begin{equation}
\begin{aligned}  
\label{l2}
\Delta 
= 
\overset{N}{\underset{\underset{i \neq k}{i=1}}{\sum}} \big( \lambda_{i}^{\Omega } - \lambda_{i \backslash \{k\}}^{\Omega } \big) \left(\overset{N}{\underset{\underset{j \neq k}{j=1}}{\sum}}  w_j^* z_{ji}  \right)^2  + \overset{n}{\underset{\underset{i \neq k}{i=1}}{\sum}} \lambda_{i \backslash \{k\}}^{\Omega }  \left\{ \left(\overset{N}{\underset{\underset{j \neq k}{j=1}}{\sum}} w_j^* z_{ji}  \right)^2 - \left(\overset{N}{\underset{\underset{j \neq k}{j=1}}{\sum}} w_{j \backslash \{k\}}^{*} z_{ji\backslash \{k\}}  \right)^2 \right\}.
\end{aligned}
\end{equation}

\bigskip

A positive value of $\Delta$ signals a larger contribution of the adjacency matrix $\mathbf{\Omega}$ to the loss function $\mathcal{Q}(\widetilde{\mathbf{\Gamma}},\mathbf{w}^*)$ than of $\mathbf{\Omega}_{\backslash \{k\}}$ to the loss function $\mathcal{Q}(\widetilde{\mathbf{\Gamma}}_{\backslash \{k\}},\mathbf{w}_{\backslash \{k\}}^*)$. In contrast, a negative value indicates a larger contribution of the reduced adjacency matrix. Intuitively, a positive $\Delta$ implies that the $(N-1)$ remaining assets are more connected in the full matrix configuration than in the reduced matrix configuration. In contrast, if $\Delta$ is negative then the $(N-1)$ assets exhibit low connectivity once we remove asset $k$ from the portfolio. Notice also that all the eigenvalues of the adjacency matrix $\mathbf{ \Omega }_{ \backslash \{k\}}$ for each $i \in \left\{ 1,..., N \right\}$ are lying in between the eigenvalues of $\mathbf{ \Omega }$ (see, \cite[p.~6]{van2014graph}).

\newpage 

This implies that  $\lambda_{i}^{\Omega } - \lambda_{i \backslash \{k\}}^{\Omega } \geq 0$. In order to obtain a better insight into this result, we compute the difference of eigenvalues between the two risk matrices. Using expression \eqref{lambda1b}, we obtain that
\begin{equation*}
\begin{aligned} 
\lambda_{s}^{\Omega } 
=  
\overset{N}{\underset{\underset{i \neq k}{i=1}}{\sum}} \overset{N}{\underset{\underset{j \neq k}{j=1}}{\sum}} z_{is} z_{js} \widetilde{ \gamma }_{(i,j)} + \overset{N}{\underset{j=1}{\sum}} z_{ks} z_{js} \widetilde{\gamma}_{(k,j)}  + \overset{N}{\underset{i=1}{\sum}} z_{is} z_{ks} \widetilde{\gamma}_{(k,i)}.
\end{aligned}
\end{equation*}
The corresponding eigenvalue of the reduced adjacency matrix  $\mathbf{\Omega}_{\backslash \{k\}}$ is  
\begin{align} \label{eigen2}
\lambda_{s \backslash \{k\}}^{\Omega } 
= 
\overset{N}{\underset{\underset{i \neq k}{i=1}}{\sum}} \overset{N}{\underset{\underset{j \neq k}{j=1}}{\sum}} z_{is\backslash \{k\}} z_{js\backslash \{k\}} \widetilde{ \gamma }_{(i,j)}.
\end{align}
Simple algebra shows that 
\begin{align} \label{diffeigen}
\lambda_{s}^{\Omega } - \lambda_{s \backslash \{k\}}^{\Omega } 
= 
\overset{N}{\underset{\underset{i \neq k}{i=1}}{\sum}} \overset{N}{\underset{\underset{j \neq k}{j=1}}{\sum}} \big( z_{is} z_{js} - z_{is\backslash \{k\}} z_{js\backslash \{k\}} \big) \widetilde{\gamma}_{(i,j)} + \overset{N}{\underset{j=1}{\sum}} z_{ks} z_{js} \widetilde{\gamma}_{(k,j)} + \overset{N}{\underset{i=1}{\sum}} z_{is} z_{ks} \widetilde{\gamma}_{(k,i)}.
\end{align} 
Expression \eqref{diffeigen} shows that the difference in eigenvalues is determined by differences in the eigenvectors of the adjacency matrices $\mathbf{ \Omega }$ and $\mathbf{ \Omega }_{\backslash \{k\}}$ and the contribution of the off-diagonal terms of the removed asset $k$. The latter two terms associated to asset $k$ are expected to be of positive sign entailing an increase in connectivity for the full portfolio compared to the reduced portfolio. There is, however, another effect given by the difference of eigenvectors between the matrices $\mathbf{Z}$ and $\mathbf{Z}_{\backslash\{k\}}$. The difference in eigenvectors can be sufficiently sizeable to offset the effect of removing node $k$, and such that expression \eqref{diffeigen} is negative. In this case $\Delta$ is likely to be negative.

As a final exercise we consider the case when the eigenvalues of the remaining $(N-1)$ assets, once asset $k$ is removed, are very similar. This is possibly the case for large portfolios. Let us assume that $\lambda_{s}^{\Omega} = \lambda_{s \backslash \{k\}}^{\Omega }$ for $s \in \left\{ 1,\ldots,k-1,k+1,\ldots,N \right\}$. Then, using \eqref{l2}, we can derive an equivalent expression for the $\Delta$ function as it is shown by Proposition \ref{Proposition5}. 

\begin{proposition}
\label{Proposition5}
Let $\Delta$ be the net difference function of the loss functions. Then, we can prove that an equivalent expression for $\Delta$ is given by the following expression
\begin{equation}
\begin{aligned}  \label{l3}
\Delta 
& =
\overset{N}{\underset{\underset{i \neq k}{i=1}}{\sum}} \lambda_{i \backslash \{k\}}^{\Omega }  \left(\overset{N}{\underset{\underset{j \neq k}{j=1}}{\sum}} \big( w_j^* + w_{j \backslash \{k\}}^{*} \big) z_{ji}  \right) \left(\overset{N}{\underset{\underset{j \neq k}{j=1}}{\sum}} \big( w_j^* - w_{j \backslash \{k\}}^* \big) z_{ji}  \right) 
\\ 
& + \overset{N}{\underset{\underset{i \neq k}{i=1}}{\sum}} \lambda_{i \backslash \{k\}}^{\Omega }  \left( \overset{N}{\underset{\underset{j \neq k}{j=1}}{\sum}} w_{j \backslash \{k\}}^{*} \big( z_{ji}+z_{ji\backslash\{k\}} \big) \right) \left(\overset{N}{\underset{\underset{j \neq k}{j=1}}{\sum}} w_{j \backslash \{k\}}^{*} \big( z_{ji}-z_{ji\backslash\{k\}} \big) \right).
\end{aligned}
\end{equation}
\end{proposition}

\newpage 

\begin{remark}
Proposition \ref{Proposition5} demonstrates that differences in connectivity between the stocks across the full portfolio vis-a-vis the reduced by one central stock portfolio are driven by differences in portfolio weights and eigenvectors of the adjacency matrices $\mathbf{ \Omega }$ and $\mathbf{ \Omega }_{\backslash\{k\}}$. 
\end{remark}
Next, we find the level of eigenvector centrality that minimizes the portfolio risk when this is expressed with respect to the adjacency matrix. 

\begin{proposition}
\label{proposition.centrality}
Let $\mathcal{Q} ( \mathbf{ \Omega }, \mathbf{w} ) = \mathbf{w}^{\prime} \mathbf{ \Omega } \mathbf{w}$ the portfolio risk function with respect to the adjacency matrix. Then, the eigenvector centrality measure which minimizes the portfolio risk is  
\begin{align}
\label{minimum.centrality}
 v_i =  - 2 w_i \bigg/  \left(\overset{N}{\underset{j=1}{\sum}} w_{j} v_{j}  \right) 
\end{align}
\end{proposition}

\begin{proof}
Notice that, by the proof of Theorem \ref{theorem1} in Appendix \ref{AppendixA} we have that
\begin{align} 
\frac{ \partial^2 \mathbf{w}^{\prime} \mathbf{ \Omega } \mathbf{w}}{  \partial^2 v_i } 
= 
\lambda^{ \Omega }_{(1)} \left(\overset{N}{\underset{j=1}{\sum}} w_{j} v_{j}  \right)^2 > 0.
\end{align}
which is always a positive quantity since we assume that $\lambda^{ \Omega }_{(1)} > 0$.
Thus, we can conclude that the eigenvector centrality which minimizes, the portfolio risk
$\mathbf{w}^{\prime} \mathbf{ \Omega } \mathbf{w}$,
is given by \eqref{minimum.centrality}.
\end{proof}
\begin{remark}
Notice that Proposition \ref{proposition.centrality} provides the optimal eigenvector centrality measure which minimizes the portfolio risk based for the GMVP strategy. However, practically obtaining the particular vector is not an easy task, which it would require to have a certain network topology which ensures that the eigenvector centrality can be expressed in that form.   
\end{remark}

Our procedure proposes a technique for dimensionality reduction in graphs. In particular, we remove central nodes based on the eigenvector centrality measure. The criterion for the removal of central stocks in our setting is determined by the $\Delta$ function as given by Proposition \ref{Proposition5}. Furthermore, we have derived an analytical expression for the eigenvector centrality given by expression \eqref{minimum.centrality} which minimizes the portfolio risk and it holds regardless we employ the portfolio risk function with respect to the risk matrix or the adjacency matrix. For instance, when the vector of eigenvector centrality has the particular form then the portfolio risk is achieved its minimum and no further node exclusions is necessary.  

In summary, via our theoretical analysis in Section \ref{SectionII} and \ref{SectionIII} we utilize the spectral vector decomposition applied to the risk matrix and the adjacency matrix as well as the eigendecomposition which corresponds to the leading eigenvalue and eigenvector to derive useful insights regarding the relationship between optimal portfolio allocation and stock centrality as well as the relationship between portfolio risk and stock centrality. We aim to illustrate that indeed the relationship between portfolio risk and the tail centrality measure holds based on an empirical application of the proposed procedure. Therefore, in Section \ref{SectionIV} we provide empirical evidence of the proposed methodology within an out-of-sample estimation framework.

\newpage 

\section{Empirical Application} 
\label{SectionIV}

The goal of our empirical study is to investigate the main research questions we tackle within our framework such as the relation between optimal portfolio allocation and stock centrality as well as portfolio risk and stock centrality using real data. These empirical finance relations are studied from the perspective of an investor with tail events concerns, therefore assessing the statistical properties of the novel risk matrix is important to verify the robustness of the proposed portfolio choice methodology. In particular, we assess the sensitivity of portfolio performance measures to the network topology using the VaR-$\Delta$CoVaR matrix. We focus on the period before and after the financial crisis of 2008 during which financial markets exhibited increased levels of financial connectedness increasing the likelihood of systemic risk effects and financial contagion across the economy (\cite{billio2012econometric}, \cite{hardle2016tenet}).

\subsection{Data Description}

In particular, our empirical study employs the dataset of  \cite{hardle2016tenet} and includes a panel of the top 100 publicly traded financial institutions by market capitalization. More precisely, these are categorized into four groups: (i) depositories, (ii) insurance companies, (iii) broker-dealers, and (iv) others; which allows us to construct a financial network and apply the proposed graph based optimal portfolio allocation methodology. The dataset contains the stock returns of these firms along with a set of macroeconomic variables both corresponding to the same period (between 5, January 2007 and 4, January 2013) based on weekly time series observations. Furthermore, the dataset includes a set of firm variables which we use as the observable firm factors. In particular, these include balance sheet information such as: (i) total assets/total equity to capture firm leverage, (ii) short term debt-cash/total liabilities to capture maturity mismatch, (iii) ratio of the market to the book value of the total equity to capture the market-to-book firm characteristics and (iv) log of the total book equity to capture the size of the firm (see, Table \ref{table 1}).

The stock returns are estimated from the historical sequences of stock prices $P_t$ across the cross-section of firms. Moreover, we transform both macroeconomic and firm variables into stationary time series before fitting to the econometric specifications, by taking first differences. On the one hand, the macroeconomic variables allow us to capture the credit and liquidity channel of systemic risk and financial interconnectedness in the network. On the other hand, the inclusion of firm variables captures their individual behavioural characteristics and reflect changes in idiosyncratic risk across time (e.g., see \cite{duffie2009frailty}). The information set is identical for both the estimation and the forecasting of the VaR-$\Delta$CoVaR risk matrix, which is suitable for an out-of-sample optimal asset allocation study. Our empirical study contributes to the literature of systemic risk monitoring and graph based optimal portfolio allocation using our proposed risk matrix constructed with tail-risk forecasts.

\newpage 

The macroeconomic state variables consist of the following set: (i) implied volatility index, VIX; (ii) short-term liquidity spread calculated as the difference between the three-month repo rate and the three-month bill rate to measure short-term liquidity risk; (iii) changes in the three-month Treasury bill rate; (iv) changes in the slope of the yield curve corresponding to the yield spread between the ten year Treasury rate and the three-month bill rate from the Federal Reserve Board; (v) changes in the credit spread between BAA rated bonds and the Treasury rate; (vi) the weekly S$\&$P500 index returns, and (vii) the weekly Dow Jones US Real Estate index returns (see, \cite{hardle2016tenet} for detailed summary statistics and Table \ref{table 1} in Appendix \ref{AppendixB} for variable definitions). For the empirical application of the procedure in our paper, we examine the use of both stationary and nonstationary data.

\subsection{Dynamic Quantile Regression Model}
\label{Section5.2}

Our proposed risk matrix belongs to the family of dynamic high dimensional covariance matrices, and captures network tail risk dependence, since is based on tail risk measures. We estimate the time-varying VaRs and CoVaRs conditional on a vector of lagged state variables $\mathbf{M}_{t-1}$. Both VaR and CoVaR risk measures represent quantiles of the distribution of returns under different conditioning sets; hence, to predict these quantities we employ the quantile regression specifications proposed by the seminal study of \cite{Koenker1978regression} (see, also \cite{KoenkerXiao02}).  

Consider the conditional quantile function estimated via $Q_{y_i} \left( \tau | x_i \right) = F_{y_i}^{-1} \left( \tau | x_i \right)$. Then, the optimization function to obtain the model estimates is expressed as below
\begin{align}
\label{quantile.estimation}
Q_{\tau} \left( y_i | x_i \right) = \underset{ q(x)}{\text{arg min}} \ \mathbb{E} \big[ \rho_{\tau} \left( y_i - q(x_i) \right) \big]
\end{align}
where $\tau \in (0,1)$ is a specific quantile level, and $\rho_{\tau}( u ) = u \left( \tau - \mathbf{1}_{ \left\{ u < 0 \right\} } \right)$ is the check function (see, \cite{newey1987asymmetric}). Suppose $1 \leq t \leq T$, then based on the  estimation method given by \eqref{quantile.estimation} the VaR and CoVaR are generated by the following econometric specifications 
\begin{align}
R_{(i),t} &= \nu_{(i)}  +  \mathbf{\xi}_{(i)}^{\prime} \mathbf{M}_{t-1} + u_{(i),t} \label{VaR} \\
R_{(j),t} &= \nu_{(j)} + \mathbf{\xi}_{(j)}^{\prime} \mathbf{M}_{t-1} + \beta_{(j)} R_{(i),t}  + v_{(j),t} \label{CoVaR}
\end{align} 
where $\{ R_{(i),t} \}$ for $i \in \left\{ 1,...,N \right\}$ is the vector of portfolio returns at time $t$, and $\mathbf{M}_{t-1}$ is a vector of exogenous regressors containing macroeconomic characteristics common across assets and $\Theta = \{ \nu_{(i)}, \mathbf{\xi}_{(i)}, \nu_{(j)}, \mathbf{ \xi }_{(j)}, \beta_{(j)} \}$ is a compact parameter space. 

Let $Q_{\tau} \left( \cdot\ | \ \mathcal{F}_{t-1} \right)$ denote the quantile operator for $\tau \in (0,1)$ conditional on an information set $\mathcal{F}_{t-1}$. Then, the error terms satisfy $Q_{\tau} \left( u_{(i),t} \ | \ \mathbf{M}_{t-1} \right)=0$ for the quantile regression model \eqref{VaR}, and $Q_{\tau} \left( v_{(j),t} \ | \ R_{(j),t}, \mathbf{M}_{t-1} \right) = 0$ for the quantile regression model given by \eqref{CoVaR}. Moreover, denote with $\theta_{(1)} = \left( \nu_{(i)}, \xi_{(i)}^{\prime} \right)$ and $\widetilde{X}_{t-1}^{(1)} = \left( 1, \mathbf{M}_{t-1}^{\prime} \right)^{\prime}$ for model \eqref{VaR}, and with $\theta_{(2)} = \left( \nu_{(j)}, \xi_{(j)}^{\prime}, \beta_{(j)} \right)$ and  $\widetilde{X}_{t-1}^{(2)} = \left( 1, \mathbf{M}_{t-1}^{\prime}, R_{(i),t} \right)^{\prime}$ for model \eqref{CoVaR}. 

\newpage 

Then, the model estimate 
\begin{align}
\widehat{ \theta}_{(1)} \left( \tau  \right) = \underset{ \theta_{(1)} \in \mathbb{R}^{p+2} }{  \text{arg min} } \sum_{t=1}^{T} \rho_{\tau} \left( R_{(i),t} - \theta_{(1)}^{\prime} \widetilde{X}_{t-1}^{(1)}  \right)
\end{align}
Similarly, the quantile estimator of \eqref{CoVaR} is obtained via the optimization function below 
\begin{align}
\widehat{ \theta }_{(2)} \left( \tau  \right) = \underset{ \theta_{(2) } \in \mathbb{R}^{p+1} }{  \text{arg min} } \sum_{t=1}^{T} \rho_{\tau} \left( R_{(j),t} - \theta_{(2)}^{\prime} \widetilde{X}_{t-1}^{(2)}  \right)
\end{align}
The state variables, that is, the macroeconomic and financial variables, included in the econometric model, capture time variation in the conditional moments of asset returns. Therefore, the VaR and CoVaR of each financial institution are estimated as vectors from a time-varying distribution. This approach allows to explain the optimal portfolio allocation in terms of changes in the systemic risk and financial connectedness in the network. To implement this, we use a rolling window, within which the quantile regressions given by  \eqref{VaR} and \eqref{CoVaR} are fitted and the $\widetilde{\mathbf{\Gamma}}$ risk matrix is constructed based on one-period ahead forecasts.   

Specifically, the one-period ahead VaR$_{t+1}$ for a coverage probability $\tau \in (0,1)$ is obtained by regressing  $R_{(i),t}$ on $\mathbf{M}_{t-1}$ as given by the first step regression (\ref{VaR}) which gives the parameter estimates $\widehat{\nu}_{(i)}$ and $\widehat{ \xi }_{(i)}$. The forecasted one-period ahead VaR is computed as below 
\begin{align} \label{equ:VaR}
\widehat{ \text{VaR} }_{(i),t+1} ( \tau ) = \widehat{\nu}_{(i)} + \widehat{ \mathbf{ \xi } }_{(i)} \mathbf{M}_{t}.  
\end{align}
Similarly, the one-period ahead forecast of CoVaR$_{t+1}$ is obtained by regressing $R_{(j),t}$ on $\mathbf{M}_{t-1}$ and $R_{(i),t}$ as shown in the second step regression (\ref{CoVaR}) using the quantile estimation as defined by \eqref{quantile.estimation}. To do this, we collect the parameter estimates $\widehat{\nu}_{(j)}, \widehat{ \xi }_{(j)}, \widehat{\beta}_{(j)}$ from the second regression, and construct the forecasted one-period ahead CoVaR measure with
\begin{align}
\label{equ:CoVaR}
\widehat{ \text{CoVaR} }_{(j),t+1} ( \tau ) = \widehat{\nu}_{(j)} + \widehat{\mathbf{ \xi  }}_{(j)}  \mathbf{M}_{t} +  \widehat{\beta}_{(j)} \widehat{VaR}_{i,t+1} ( \tau ),
\end{align}
where $\widehat{VaR}_{i,t+1}( \tau )$ replaces the actual $VaR_{i,t+1} ( \tau )$ measure.
Then, the $\Delta \text{CoVaR}$ measure is computed as the difference of the two risk measures
\begin{equation}
\Delta \text{CoVaR} = \widehat{\text{CoVaR}}_{(j),t+1} ( \tau ) - \widehat{\text{VaR}}_{(j),t+1 } ( \tau ),
\end{equation}
An important aspect of our proposed estimation methodology is the assumption of a graph representation with nodes being the stock returns and edges being pairwise tail forecasts, which can be thought as the level of simultaneous Granger causality due to the existence of tail and graph dependence. Therefore, these tail forecasts are estimated using the aforementioned predictive regression models based on the bipartite graph structure, which implies an estimation implementation for each pair $(i,j)$ of nodes for the graph $\mathcal{G} = (\mathcal{V}, \mathcal{E})$. 

\newpage 

The construction of one-period ahead forecasts for the dynamic portfolio allocation, based on a graph representation, consists one of the main contributions of the empirical application of the paper. The particular estimation procedure is employed for each of the pairwise predictive regressions, to obtain the elements of the VaR-$\Delta$CoVaR matrix. Next, we examine  the model fit and specification of the proposed risk matrix $\widetilde{ \mathbf{\Gamma} }$.

\subsection{Analysis of Model Fit and Specification}

In Section \ref{Section5.2} we described the procedure for jointly estimating and forecasting the VaR and CoVaR for the financial institutions in the network. The estimation procedure is straightforward to implement, however there is now a growing statistical literature which studies the statistical conditions for joint elicitability of these risk measures, that permit to employ the aforementioned econometric specifications. Related studies include \cite{ziegel2016coherence}, \cite{fissler2016higher}, \cite{Patton2019dynamic} and  \cite{frongillo2020elicitation}. Therefore, here we implement various robustness checks focusing on the out-of-sample accuracy of the procedure, and the robustness of the model fit via alternative specifications of the quantile regression models, using formal statistical tools such as quantile specification tests (see \cite{yang2021semiparametric}).  

\subsubsection{Factors: Firm Characteristics}

Our starting point for the model fit evaluation, is to account for the effects of unobserved heterogeneity due to differences under the presence of latent factors such as firm specific variables. To do this, we examine the goodness-of-fit for the quantile risk measures in the case of no firm characteristics vis-a-vis the case with the additional observable firm characteristics as predictors. In particular, we  evaluate the performance of the quantile models when the firm variables of firm $i$ are included in the specification of the two quantile predictive regressions corresponding to the VaR and CoVaR risk measures. The purpose of distinguishing this particular aspect is to reduce any potential biased tail forecasts due to possible misspecified pairwise predictive models. Furthermore, we can examine whether our theoretical results are still valid in the case we obtain the optimal asset allocation conditional on firm characteristics. 

The addition of firm variables can provide additional predictive accuracy in the out-of-sample estimation of the risk matrix resulting to improve accuracy when obtaining an estimate for the sharpe ratio. On the other hand, considering firm characteristics to the optimal portfolio choice problem we contribute to the related literature which is in favour of incorporating such variables (see, \cite{mcgee2020optimal}). More precisely, the particular approach of constructing the optimal portfolio choice problem using firm characteristics is quite recent and provides a methodology for reduced model uncertainty when fitting asset pricing models. Specifically, using data on firm factors can employed to reduce the uncertainty induced in the estimation procedure which occurs from the presence of any further unobservable factors.

\newpage 

Therefore, since we consider the optimal portfolio choice problem in graphs the inclusion of firm characteristics can also provide a tractable way of improving the information set and as a result modelling more formally the network tail dependence induced by the underline node specific distributions. This is certainly an aspect which has not been examined before in the literature of portfolio choice problems under graph dependence.  Furthermore, when obtaining the optimal number of stocks based on eigenvector centrality, capturing correctly the tail connectivity of stocks by the corresponding centrality measure, robustifies the estimation of the eigendecomposition and reduces the likelihood of obtaining a low-rank risk matrix.

\subsection{Dynamic Optimal Portfolio Allocation under Tail Risk}

The dynamic optimal portfolio allocation for an investor concerned with tail risk can be constructed using the VaR-$\Delta$CoVaR matrix. We implement the dynamic optimization with a rolling window forecasting scheme of a fixed length. Within this setting, the investor's objective function is expressed as below
\begin{align}
 \mathbf{w}_{t}^{*} =  \underset{\mathbf{w}_{t} \in \mathbb{R}^N }{\text{arg min}} \big\{ \mathbf{w}^{\top}_{t} \widetilde{ \mathbf{\Gamma} }_{t+1|t} \mathbf{w}_{t} \big\} \ \ \textit{subject to }  \ \ \ \mathbf{w}^{\top}_{t} \mathbf{1}= 1, \label{eq1w}
\end{align}
where the dynamic risk matrix $\widetilde{ \mathbf{\Gamma} }_{t+1|t}$ replaces the static matrix $\widetilde{ \mathbf{\Gamma} }$ discussed in \eqref{gammamat}.

The full sample contains $314$ weekly observations spanning the period between 5, January 2007 and 4, January 2013, covering a period before and after the 2008 financial crisis. Due to the rapid development of the events that led to the market crashdown in September 2008, the weekly frequency is a suitable sampling frequency to capture the dynamic systemic risk and increased interconnectedness as documented in the literature (see, \cite{hardle2016tenet}). 

The forecasting scheme which is based on a rolling window with sample size $\kappa=250$ time series observations, spans the period between 5, January 2007 and 14, October 2011. We find this sampling period to provide a good approximation of a highly connected network topology, that allows to apply our theoretical framework. Using this sample, we obtain a forecast of the risk matrix $\widetilde{ \mathbf{\Gamma} }_{t+1|t}$ corresponding to 21 October 2011, and construct the corresponding optimal portfolio weights. This procedure is repeated until 03 January 2013 such that we reestimate the risk matrix weekly using the rolling window and compute the vector of optimal portfolio weights in each period. The out-of-sample distribution of portfolio returns is calculated as $r_{t+1}^p = \mathbf{w}^{\prime}_{t} \mathbf{R}_{t+1}$, where $\mathbf{w}^{\prime}_{t}$ is the estimated vector of optimal weights and $\mathbf{R}_{t+1}$ the stock return of the assets included in the portfolio at time $t+1$.

\subsubsection{Portfolio performance measures}

Portfolio performance is assessed using standard performance metrics. We consider the out-of-sample Sharpe Ratio obtained over the out-of-sample evaluation period:
\begin{align}
\widehat{SR} = \frac{ \bar{R}_p }{ \widehat{\sigma}_{R_p} }
\end{align}
where $\bar{R}_p$ and $\widehat{\sigma}_{R_p}$ are the out-of-sample mean and variance of the returns. Thus,
\begin{align}
 \bar{R}_p  = \frac{1}{N - \kappa} \sum_{t = \kappa+1}^{T} r_{t}^p \ \ \text{and} \ \ \widehat{\sigma}^2_{R_p} = \frac{1}{N - \kappa -1} \sum_{t = \kappa+1 }^{T} (r_{t}^p  - \bar{R})^2.
\end{align}

\medskip

\subsubsection{Optimal number of stocks based on eigenvector centrality}

The next aspect we examine is the determination of a possible optimal number of stocks based on the eigenvector centrality measure and utilizing a sequential procedure for excluding central nodes from the graph, as described by Algorithm \ref{alg1}. Firstly, we apply the dynamic optimal portfolio problem for all rolling windows of the full sample with 314 weekly observations and $\kappa=250$ being the length of each rolling window as explained in details in the previous section. Secondly, we investigate the sensitivity of the portfolio performance measures to the network topology which allows to investigate empirically the relationship between stock centrality and portfolio risk. In particular, we examine the effect of network topology on the portfolio mean and variance under based on the proposed risk function which accommodates tail connectivity. To do this, we design a procedure in which in each rolling window, we modify the network topology (by excluding central nodes). Thus, the particular sequential procedure based on the portfolio choice closed form solution is applied, after removing the most central asset from the network estimated in each replication step. 

Furthermore, notice that our procedure chooses the optimal number of stocks based on eigenvector centrality since we propose a graph-based portfolio selection mechanism by being agnostic about the industry specific characteristics of stocks, which would require a different construction of the portfolio investment strategy. For instance, the framework of \cite{chen2020application} propose a procedure for optimal portfolio choice problems using sparse-group lasso regularization techniques allowing for both asset allocation and sector selection. A different approach is to follow an efficient sorting algorithmic procedure for obtaining the optimal asset allocation as in the framework proposed by \cite{ledoit2019efficient}. Therefore, our framework consider tail connectivity across the nodes taking into consideration the network topology rather than any clustering effects among stocks.

\newpage

In summary, to evaluate the sensitivity of assets to the network topology in the tails in each rolling window, we estimate the VaR-$\Delta$CoVaR matrix for the full sample of $N=100$ assets and then re-estimate the risk matrix to derive the optimal number of assets in the portfolio for each period by applying Algorithm 1\footnote{We ensure that the optimal weight vector is always positive using the constrained optimization routine proposed in the R package of \cite{scrucca2013ga}. The particular package in R provides a set of Genetic Optimization Algorithms commonly used in portfolio optimization and other linear programming settings. }. Our proposed Algorithm determines the optimal number of assets in the portfolio. The procedure is based on eigenvector centrality and excludes central stocks from the portfolio based using the criterion function $\Delta$ that checks the difference for the corresponding quadratic loss functions before and after one additional exclusion\footnote{For the interested reader  Algorithm 1 in \cite{calvo2021granger} proposes a procedure for excluding nodes within a neural network which can increase the predictive accuracy within a high dimensional setting.} of a node from the network.  

\begin{algorithm} \label{alg1}

Let $\mathbf{r}_t^p=\overset{n}{\underset{i=1}{\sum}} w_i^* R_{i,t}$ be a portfolio with associated risk matrix $\widetilde{\mathbf{\Gamma}}$, where $\textbf{w}^*$ is the solution to the objective function \eqref{minVaR}.
\begin{enumerate}
\item[Step 1.] Solve the optimization problem for this portfolio and obtain
\begin{equation}
\mathbf{w}^{*} = \frac{\widetilde{\mathbf{\Gamma}}^{-1} \mathbf{1}}{\mathbf{1}^{\prime} \widetilde{\mathbf{\Gamma}}^{-1} \mathbf{1}}.
\end{equation}

\item[Step 2.] Construct the adjacency matrix $\mathbf{\Omega}$, obtain the eigenvalues and let $\lambda=\max(\lambda_1,\ldots,\lambda_N)$. The eigenvalue centrality measure \eqref{central} is
\begin{align}
v_i = \lambda^{-1} \overset{n}{\underset{j=1}{\sum}} \Omega_{ij} \hspace{0.2ex} v_j.
\end{align}

\item[Step 3.] Let $v_k = \text{max}(v_1,\ldots,v_n)$. Then, asset $k$ exhibits the largest eigenvector centrality statistic. We construct the reduced portfolio $\mathbf{r}_{t \backslash {(k)}}^p=\overset{N}{\underset{\underset{i \neq k}{i=1}}{\sum}} \mathbf{w}_{i \backslash \{k\}}^* R_{i,t}$ obtained from removing asset $k$, and obtain the associated risk matrix $\widetilde{\mathbf{\Gamma}}_{\backslash \{k\}}$.

\item[Step 4.] Solve the optimization problem \eqref{minVaR} for this portfolio and obtain
\begin{equation}
\mathbf{w}_{\backslash \{k\}}^{*} = \frac{\widetilde{\mathbf{\Gamma}}_{\backslash \{k\}}^{-1} \mathbf{1}}{\mathbf{1}^{\prime} \widetilde{\mathbf{\Gamma}}_{\backslash \{k\}}^{-1} \mathbf{1}}.
\end{equation}

\item[Step 5.] Compute the condition given by the function $\Delta$. Remove asset $k$ if this condition is negative. The Alogirthm stops the first time $\Delta$ becomes positive.

\item[Step 6.] If the asset is removed repeat the procedure from Step 2 for the portfolio with $N-1$ assets. Otherwise, we retain the portfolio with $N$ assets.
\end{enumerate}
\end{algorithm}

\newpage 

\subsection{Empirical results}

To begin with, for the empirical application of this paper we focus on the quantile level of $\tau = 5\%$, which is a commonly used level of risk at the tails of the underline distributions considered for financial risk management purposes. Moreover, we consider the minimum-variance portfolio strategy for the full sample of all financial institutions as well as for the cases of the low versus high network inteconnectedness. In particular, to represent these two scenarios we consider removing the most central nodes from the financial network for the first scenario, while we exclude the nodes with low centrality in the second scenario, based on the eigenvalue centrality. As a robustness check of our proposed methodology we also use alternative centrality measures. Our empirical results indicate that the network topology has a direct impact on the portfolio performance when an investor is concerned with tail events. As we demonstrate on Table \ref{table_main_results} below, when testing for differences in the Sharpe ratios of the two network topologies that correspond to the high versus the low interconnectedness, the null hypothesis of no differences is rejected at significance level 0.05 with bootstrap p-values indistinguishable from zero. To provide evidence of statistical difference of the distribution of portfolio returns from the different network topologies we apply the Sharpe ratio test of \cite{ledoit2008robust} (see, the R package of \cite{ardia2017package}).  

\begin{table}[h!]
  \centering
  \caption{\\ \textbf{LW robust hypothesis testing for the Sharpe ratio} }
    \begin{tabular}{c|cc|cc|cc}
    \hline
    \hline
          & X     & Y     & X     & Z     & Y     & Z \\
    \hline
    Sharpe ratios & 0.1461 & 0.2039 & 0.1461 & 0.2076  & 0.2039 & 0.2076 \\
    Sharpe ratios difference ($\widehat{\Delta}$) & -0.0578 &       & -0.0615 &       & -0.0037 &  \\
    \hline
    Asymptotic & \multicolumn{6}{c}{} \\
    \hline
    t-Statistic & -1.9984 & \multirow{2}[0]{*}{} & -1.8545 & \multirow{2}[0]{*}{} & -0.2810 & \multirow{2}[0]{*}{} \\
    p-value & 0.0457$^{**}$ &       & 0.0637 &       & 0.7787 &  \\
    \hline 
     Asymptotic (HAC) & \multicolumn{6}{c}{} \\
    \hline
    t-Statistic & -1.8852 & \multirow{2}[0]{*}{} & -1.7027 & \multirow{2}[0]{*}{} & -0.2687 & \multirow{2}[0]{*}{} \\
    p-value & 0.0594 &       & 0.0886 &       & 0.7881 &  \\
    \hline
    \textit{i.i.d} Bootstrap  & \multicolumn{6}{c}{} \\
    \hline
    t-Statistic & -1.9984 & \multirow{2}[0]{*}{} & -1.8545 & \multirow{2}[0]{*}{} & -0.2810 & \multirow{2}[0]{*}{} \\
    p-value & 0.0484$^{**}$ &       & 0.0686 &       & 0.7886 &  \\
    \hline
    Circular Bootstrap  & \multicolumn{6}{c}{} \\
    \hline
    t-Statistic & -2.2534 & \multirow{2}[0]{*}{} & -2.0923 & \multirow{2}[0]{*}{} & -0.3452 & \multirow{2}[0]{*}{} \\
    p-value & 0.0480$^{**}$ &       & 0.0639 &       & 0.7872 &  \\
    \hline
    \hline
    \end{tabular}%
   \label{table_main_results}%
\end{table}%
\begin{small}
Table \ref{table_main_results} corresponds to an out-of-sample estimation window of size $T = 250$ with $n = 64$.  Both the \textit{i.i.d} and the circular Bootstrap are computed based on $B = 10,000$ replications. Statistical significance for the tests is based on estimated p-values and indicated with $(^{*})$ for $\alpha = 10 \%$, $(^{**})$ for $\alpha = 5 \%$ and $(^{***})$ for $\alpha = 1 \%$.
\end{small}

\newpage 

Our empirical results as summarized on Table \ref{table_main_results} demonstrate the Sharpe ratio test proposed by \cite{ledoit2008robust}) for robust portfolio performance evaluation, applied to an out-of-sample estimation window with 250 time observations. We observe that both the asymptotic t-statistic as well as the iid and circular bootstrapped t-statistics indicate that statistical significant evidence of differences in the Sharpe ratios of two distributions of portfolio returns which correspond to the network topology of high interconnectedness versus low interconnectedness respectively, that is, the pair $\left\{ X_j, Y_j \right\}_{j=1}^{T_0}$ above. Furthermore, when we compare each of these generated portfolio return distributions separately against the portfolio return distribution of the full sample we find no statistical evidence of differences between the two. The latter verifies that by having nodes of high centrality in the network mixed with nodes of low centrality is practically equivalent to the full sample. Therefore statistical significant differences appear only in the case we compare the portfolio return distributions after trimming the most central assets versus the case of trimming the low centrality assets in the network. Notice that these results correspond to the macroeconomic covariates after taking the first difference which ensures stationary time series properties.

In summary Tabel \ref{table_main_results} above demonstrates that the portfolio where we remove the most central assets Y and the portfolio with all assets Z are the best performing portfolios compared to X. However, there are no statistical differences between the portfolios Y and Z in terms of Sharpe ratio as indicated by the large p-value. In contrast the portfolio where we remove the low centrality assets, inducing a high centrality topology, is outperformed by the low centrality network topology. Therefore, based on these results it may be reasonable to conjugate that the distribution of portfolio returns that corresponds to high centrality is outperformed from the distribution of portfolio returns that corresponds to a low centrality topology. In other words, the low centrality assets have an important role for portfolio diversification. This finding constitutes one of the main theoretical and empirical contributions of this paper.

\subsubsection{Bootstrap Sharpe Ratio Testing}  

To provide statistical evidence of differences among the estimated distributions of portfolio returns that corresponds to the high  centrality versus low centrality network topologies we utilize the Sharpe ratio test of \cite{ledoit2008robust}. Assume that we have two investment strategies $A$ and $B$ whose excess returns over a given benchmark at time $t$ are $r_{tA}$ and $r_{tB}$, respectively. For instance, this could be the risk-free rate  however within our framework, we consider the benchmark to be the out-of-sample distribution of portfolio returns that corresponds to the the full sample, that is, when no nodes have been excluded within each moving window. Then, since our aim is to compare the distribution of portfolio returns of the two network topologies, we denote with $r_{tA}$ and $r_{tB}$ the network topology with high interconnectedness and the network topology with the low interconnectedness respectively and follow the bootstrap procedure proposed by \cite{ledoit2008robust}.

\newpage 

\section{Simulation Evidence}

In this Section, we provide simulation evidence to support the theoretical and empirical findings of the paper. To simulate both stock returns and economic shocks via the presence of exogenous macroeconomic variables we consider a data generating process which is standard in the stock predictability literature\footnote{Notice that we avoid to simulate a factor return model (such as in \cite{maillet2015glob}) which commonly used for estimating asset pricing models due to the structure of the VaR-$\Delta$CoVaR risk matrix.}. After obtaining a vector with the simulated data, we fit the econometric models for the VaR and CoVaR and then obtain an estimate for the risk matrix $\widetilde{\mathbf{\Gamma}}$, based on the simulated data. 

Notice that Corollary \ref{corollary1} provides theoretical evidence that the portfolio risk defined by $\mathcal{Q}(\widetilde{\mathbf{\Gamma}},\mathbf{w})$ is increasing on the centrality measure $v_i$. In practise, this conjecture implies that the higher the centrality vector the higher is the expected value of the portfolio risk as measured by the tail risk matrix (not the volatility matrix), therefore is an empirical finance result  which corresponds to tail dependent events. Therefore, when we remove central nodes from the graph based on the eigenvector centrality vector we suppose that the reduced graph has low interconnectivity and therefore the tail dependence is kept at low levels. However, this fact does not necessarily provide evidence regarding the effect of the centrality vector induced from the full model in comparison to the centrality vector induced by the reduced model after excluding one central node. Clearly the particular issue requires further investigation and a more sophisticated algorithm which is designed to select those combinations of nodes based on their centrality in relation to expected portfolio return and risk.

Furthermore, Table \ref{table:sim} provides simulation evidence about the performance of the portfolio mean, portfolio risk as well as the $\Delta$ function when excluding sequentially one central stock at a time, based on the eigenvector centrality measure for an in-sample exercise. 

\begin{table}[h!]
  \centering
  \caption{\textbf{Simulation Evidence}}
    \begin{tabular}{c|ccc}
    \hline
    \hline
    \textbf{Exclusions} & $r^p$ & $\mathcal{Q}(\widetilde{\mathbf{\Gamma}},\mathbf{w})$ & $\Delta$ \\
    \hline
    1     & 3.202 & 2.268 & -5.475 \\
    2     & 7.511 & 3.636 & 1.445 \\
    3     & 8.316 & 3.683 & -0.145 \\
    4     & 9.264 & 4.066 & -0.891 \\
    5     & 13.105 & 6.567 & -5.277 \\
    6     & 11.988 & 8.007 & -4.956 \\
    7     & 14.804 & 8.406 & -0.929 \\
    8     & 14.476 & 8.499 & -0.208 \\
    9     & 13.553 & 9.264 & -1.788 \\
    10    & 13.950 & 9.420 & -0.816 \\
    \hline
    \hline
    \end{tabular}%
  \label{table:sim}%
\end{table}%
    
\begin{small}
Table \ref{table:sim} corresponds to a data generating process with $N = 40$ nodes, $k = 7$ predictors and $t \in \left\{ 1,..., T \right\}$ where $T = 500$. 
\end{small}

\newpage 

\section{Conclusion} 
\label{SectionV}

In this study we have proposed a new risk matrix given by the VaRs and $\Delta \text{CoVaR}$ of the assets in a portfolio which can be considered as an alternative way of parametrizing volatility under uncertainty. VaR measures replace the variance and $(\text{VaR}_{i,t} \text{VaR}_{j,t})^{1/2} \Delta \text{CoVaR}_{i|j,t}$ replace the covariance terms. This risk matrix aims to capture tail events given by idiosyncratic tail risks and connectivity between assets in the tails. In this way, we focus on higher-order dependence structures across the distributions of stock returns. Furthermore, we derive the conditions under which the corresponding quadratic form has a single closed-form solution. This solution is interpreted as the optimal portfolio allocation of an investor that minimizes a loss function given by the quadratic form associated to the novel risk matrix. The particular, optimal vector of weights is similar in spirit to \cite{Markowitz52}'s minimum variance and mean-variance optimal portfolios. The proposed optimal portfolio allocation procedure can be helpful when removing stocks iteratively from a portfolio in order to obtain efficient asset allocations by taking into consideration the stock centrality.


\clearpage

\paragraph{\Large{ Appendix} }

\appendix

\section{Technical Proofs}
\label{AppendixA}

\noindent \textbf{Proof of Proposition \ref{proposition4}:}

\begin{proof}
We deduce from expression \eqref{lambda1b} that the eigenvalue $\lambda_{k}^{ \Omega }$ is increasing on the magnitude of the tail connectivity measure $\gamma_{(i,j)}$ if and only if $z_{ik} z_{jk}>0$ for each $i,j \in \left\{ 1,\ldots,N \right\}$. The proof of this result is immediate by noting that the first derivative of $\lambda_{k}^{ \Omega }$ with respect to $\gamma_{(i,j)}$ is positive if and only if $z_{ik} z_{jk}>0$ for each $i,j \in \left\{ 1,\ldots,N \right\}$. To show this, note that 
\begin{align}
\frac{\partial \lambda_{k}^{ \Omega } }{\partial \gamma_{(i,j)} } = \frac{\partial \lambda_{k}^{ \Omega } }{\partial \widetilde{\gamma}_{(i,j)} } \frac{\partial \widetilde{\gamma}_{(i,j)} }{\partial \gamma_{(i,j)} } + \frac{\partial \lambda_{k}^{ \Omega } }{\partial \widetilde{\gamma}_{(j,i)}} \frac{\partial \widetilde{\gamma}_{(j,i)}}{\partial \gamma_{(j,i)} }  
\end{align}
given that $\widetilde{\gamma}_{(j,i)} = \widetilde{\gamma}_{(i,j)}$ by construction of the risk matrix.

Note also that $\frac{\partial \lambda_{k}^{ \Omega } }{\partial \widetilde{ \gamma }_{(i,j)}} =  z_{ik} z_{jk}$ and $\frac{\partial \widetilde{ \gamma }_{ (i,j) } }{\partial \gamma_{(i,j)} }= \frac{\partial \widetilde{ \gamma }_{(j,i)} }{\partial \gamma_{(j,i)} } = \frac{1}{2}$. Then, $\frac{\partial \lambda_{k}^{ \Omega } }{\partial \gamma_{(i,j)} }=z_{ik} z_{jk}$, that is positive under the condition that $z_{ik} z_{jk}>0$ for all $i,j \in \left\{ 1,\ldots,N \right\}$. Now, using the above definitions, the quadratic form $\mathbf{w}^{\prime} \mathbf{ \Omega } \mathbf{w}$ satisfies the following expression 
\begin{align} 
\label{lossgamma.appendix}
\mathcal{Q} ( \mathbf{ \Omega } , \mathbf{w} ) := \mathbf{w}^{\prime} \mathbf{ \Omega } \mathbf{w} =   \mathbf{w}^{\prime} \left( \mathbf{Z}_{ \Omega} \mathbf{D}_{ \Omega } \mathbf{Z}_{ \Omega }^{\prime} \right) \mathbf{w} = \overset{N}{ \underset{k=1}{\sum}} \lambda_k^{ \Omega }  \left( \overset{N}{\underset{j=1}{\sum}}  w_j z_{jk}  \right)^2,
\end{align}
such that
\begin{align}
\frac{\partial \mathbf{w}^{ \prime } \mathbf{ \Omega } \mathbf{w} }{\partial \gamma_{(i,j)} } = \overset{N}{ \underset{k=1}{\sum}} \frac{\partial \lambda_k^{ \Omega } }{\partial \gamma_{ (i,j) } }  \left(\overset{N}{\underset{j=1}{\sum}} w_j z_{jk}  \right)^2 =  \overset{N}{ \underset{k=1}{\sum}} z_{ik} z_{jk} \left(\overset{N}{\underset{j=1}{\sum}} w_j z_{jk}  \right)^2. 
\end{align}
Then, under the condition $z_{ik} z_{jk}>0$ it follows that $\frac{\partial \mathbf{w}^{\prime} \mathbf{ \Omega } \mathbf{w}}{\partial \gamma_{ (i,j) } }$ is strictly positive, hence, the quadratic form induced by the adjacency matrix $\mathbf{ \Omega }$ is increasing on the risk measure $\gamma_{(i,j)}$.
\end{proof}

\bigskip

\noindent \textbf{Proof of Theorem \ref{theorem1}:}

\begin{proof}
To show the result proposed by Theorem \ref{theorem1} note that from expression \eqref{decom} 
\begin{align}
\label{decomp1}
\mathcal{Q} ( \widetilde{ \mathbf{ \Gamma }} , \mathbf{w} ) 
= 
\left\{ \mathbf{w}^{ \prime } \mathbf{ \Omega } \mathbf{w} + \sum_{i=1}^{N} w_{i}^2 \gamma_i  \right\}, 
\end{align}
since  $\mathbf{ \Omega } := \widetilde{ \mathbf{ \Gamma }} - \text{diag}( \widetilde{ \mathbf{ \Gamma }} )$. Then, the first term of \eqref{decomp1} can be expressed as below
\begin{align}
\mathbf{w}^{\prime} \mathbf{ \Omega } \mathbf{w} 
= 
\sum_{i=1}^{N} \lambda^{ \Omega }_{i} \left( \sum_{j=1}^{N} w_j z_{ji}  \right)^2.
\end{align}
such that the decomposition for $\mathbf{w}^{\prime} \mathbf{ \Omega } \mathbf{w}$ follows from Proposition \ref{proposition4}.  

\newpage 

Notice that the term $w_{i}^2 \gamma_i \neq 0$ because $\gamma_i$ represents the diagonal elements of the risk matrix, that is, the VaR of firms. Then, we obtain that 
\begin{align} 
\label{proploss2}
\mathcal{Q} ( \mathbf{ \Omega } , \mathbf{w} ) := \mathbf{w}^{\prime} \mathbf{ \Omega } \mathbf{w} 
= 
\lambda^{ \Omega }_{(1)} \left( \sum_{j=1}^{N} w_{j} v_{j}  \right)^2 + \sum_{i=2}^{N} \lambda^{ \Omega }_{i} \left( \sum_{j=1}^{N} w_j z_{ji}  \right)^2.
\end{align}

where $\lambda^{ \Omega }_{(1)}$ is the Frobenius eigenvalue. Therefore,  expression \eqref{proploss2} follows by utilizing the notation for the eigenvector that corresponds to the  Frobenius eigenvalue, which is the centrality measure $v_i$. Then, we consider the first derivative of this loss function with respect to $v_i$ we obtain the following expression
\begin{align}
\frac{ \partial \mathbf{w}^{ \prime } \mathbf{ \Omega } \mathbf{w}}{  \partial v_i } = \frac{ \partial \lambda^{ \Omega }_{(1)} }{  \partial v_i } \left( \sum_{j=1}^{N} w_{j} v_{j}  \right)^2 + 2 \lambda_{ (1) }^{ \Omega } w_i \left( \sum_{j=1}^{N} w_{j} v_{j} \right).
\end{align}

From the decomposition in \eqref{lambda1b} we have that $\lambda^{ \Omega }_{(1)} =\overset{N}{\underset{i=1}{\sum}} \overset{N}{\underset{\underset{j \neq i}{j=1}}{\sum}} v_{i} v_{j} \widetilde{ \gamma }_{i|j}$. By construction of the adjacency matrix we have that $\widetilde{ \gamma }_{i|j} = \Omega_{ij}$ for $i \neq j$ and such that the first derivative of the eigenvalue $\lambda^{ \Omega }_{(1)}$ \textit{w.r.t} $v_i$ satisfies that $\frac{ \partial \lambda^{ \Omega }_{(1)} }{  \partial v_i } = \overset{N}{\underset{\underset{j \neq i}{j=1}}{\sum}} \Omega_{ij} v_{j}$. By definition of eigenvector centrality in \eqref{central0}, we have $\overset{N}{\underset{\underset{j \neq i}{j=1}}{\sum}} \Omega_{ij}v_{j} = \lambda^{ \Omega }_{(1)} v_i$, such that
\begin{align} 
\label{derlambda}
\frac{ \partial \mathbf{w}^{\prime} \mathbf{ \Omega } \mathbf{w}}{  \partial v_i } 
= 
\lambda^{ \Omega }_{(1)} v_i \left(\overset{N}{\underset{j=1}{\sum}} w_{j} v_{j}  \right)^2 + 2 \lambda_{(1)}^{ \Omega } w_i \left(\overset{N}{\underset{j=1}{\sum}} w_{j} v_{j}  \right).
\end{align}
Notice that as we explain above the term $\frac{ \partial \lambda^{ \Omega }_{(1)} }{  \partial v_i }$ is equivalent to $\lambda^{ \Omega }_{(1)} v_i$. Therefore, for the analytical expression given by \eqref{derlambda} we can conclude that the first derivative of $\mathbf{w}^{\prime} \mathbf{ \Omega } \mathbf{w}$ with respect to $v_i$ can be positive or negative depending on the combination of the magnitude and size of the  quantities $\lambda^{ \Omega }_{(1)}$, $w_i$ and $v_i$ for $i= \left\{ 1,\ldots,N \right\}$ in the above expression. Hence, in general, no monotonic relationship prevails between asset centrality and the quadratic form $\mathbf{w}^{ \prime } \mathbf{ \Omega } \mathbf{w}$. Moreover, since $ \frac{\partial}{ \partial v_i } \sum_{i=1}^{N} w_{i}^2 \gamma_i =0$ for all $i \in \left\{ 1,..., N \right\}$, then we can determine whether there exists an increasing or decreasing relation of the portfolio risk $\mathcal{Q} ( \widetilde{ \mathbf{ \Gamma }} , \mathbf{w} )$ w.r.t to the centrality measure $v_i$ by determining the behaviour of $\mathbf{w}^{ \prime } \mathbf{ \Omega } \mathbf{w}$.  

Thus, we can only conclude that a monotonic (i.e., increasing) relation exists between the portfolio risk $\mathcal{Q}(\widetilde{\mathbf{\Gamma}},\mathbf{w})$ and stock centrality, measured by the eigenvector centrality, if and only if the quantity 
$\left\{ v_i \left(\overset{N}{\underset{j=1}{\sum}} w_{j} v_{j} \right)^2 + 2 w_i \left(\overset{N}{\underset{j=1}{\sum}} w_{j} v_{j}  \right) \right\} > 0$, which is indeed always positive since we assume that $w_i > 0$ and $v_i > 0$ $\forall \ i \ \in \left\{ 1,...,N  \right\}$. 
\end{proof}

\newpage

\noindent \textbf{Proof of Proposition \ref{proposition7}:}

\begin{proof}
We use the closed-form solution proved by Proposition \ref{propositon3} given as below
\begin{align}
\mathbf{w}^{*} = \frac{\widetilde{\mathbf{\Gamma}}^{-1} \mathbf{1}}{\mathbf{1}^{\prime} \widetilde{\mathbf{\Gamma}}^{-1} \mathbf{1} }
\end{align}
To begin with, we express the matrix $\widetilde{\mathbf{\Gamma}}$  as the sum of the following matrices 
\begin{align}
\widetilde{\mathbf{\Gamma}} \equiv \left[ \mathbf{I}_N - \left\{ \left(  \mathbf{I}_N - diag(\widetilde{\mathbf{\Gamma}} ) \right) + \left( diag ( \widetilde{\mathbf{\Gamma}} ) - \widetilde{\mathbf{\Gamma}} \right) \right\} \right]
\end{align}
We denote 
\begin{align}
\widetilde{\mathbf{\Gamma}} \equiv \big( \mathbf{I}_N - \widetilde{ \mathbf{ \Omega }} \big)
\end{align}
where $\widetilde{ \mathbf{ \Omega }} = - \left( \mathbf{B} + \mathbf{ \Omega } \right)$ such that $\mathbf{B} = \left( diag(\widetilde{\mathbf{\Gamma}}) - \mathbf{I}_N \right)$ and $ \mathbf{ \Omega } = \left( \widetilde{\mathbf{\Gamma}} - \text{diag}( \widetilde{\mathbf{\Gamma}} ) \right)$. 

Furthermore, 
\begin{equation}
\widetilde{ \mathbf{\Gamma} }^{-1}= \left( \mathbf{I}_N - \widetilde{ \mathbf{ \Omega }} \right)^{-1} = \overset{\infty}{\underset{ j = 0 }{\sum}} \left( \widetilde{ \mathbf{ \Omega }} \right)^{j}.
\end{equation}
for a sequence of real-valued matrices $\widetilde{ \mathbf{ \Omega }} \in \mathbb{R}^{N \times N}$. We have also used previously the SVD of the $\widetilde{ \mathbf{ \Omega }}$ matrix, such that $\widetilde{ \mathbf{ \Omega }} = \mathbf{S} \mathbf{D}_{ \widetilde{ \mathbf{\Gamma} } } \mathbf{S}^{\prime}$. Therefore, we obtain 
\begin{align}
\widetilde{ \mathbf{\Gamma} }^{-1} = \left( \mathbf{I}_N - \widetilde{ \mathbf{ \Omega }} \right)^{-1} = \overset{\infty}{\underset{ j = 0 }{\sum}} \left( \widetilde{ \mathbf{ \Omega }} \right)^{j} = \overset{\infty}{\underset{ j = 0 }{\sum}}    \left( \mathbf{S} \mathbf{D}_{ \widetilde{ \mathbf{ \Omega }} } \mathbf{S}^{\prime} \right)^{j} 
\end{align}
Notice that due to the orthogonal eigenvectors  $\mathbf{S}$ and $\mathbf{S}^{\prime}$, when the SVD is raised to a power $j$ we obtain the following result\footnote{For example, $\widetilde{ \mathbf{ \Omega }}^2  = \left( \mathbf{S} \mathbf{D}_{ \widetilde{ \mathbf{ \Omega }} } \mathbf{S}^{\prime} \right) \left( \mathbf{S} \mathbf{D}_{ \widetilde{ \mathbf{ \Omega }} } \mathbf{S}^{\prime} \right) = \left( \mathbf{S} \mathbf{D}^2_{ \widetilde{ \mathbf{ \Omega }} } \mathbf{S}^{\prime} \right)$. In a similar manner, we can show that this result holds for an infinite sum, due to the fact that $\mathbf{S}^{\prime} \mathbf{S} = \mathbf{I}$. }, since $\mathbf{S}^{\prime} \mathbf{S} = \mathbf{I}$,  
\begin{align}
\overset{\infty}{\underset{ j = 0 }{\sum}} \left( \mathbf{S} \mathbf{D}_{ \widetilde{ \mathbf{ \Omega }} } \mathbf{S}^{\prime}  \right)^{j}  = \mathbf{S} \left\{ \overset{\infty}{\underset{ j = 0 }{\sum}}  \mathbf{D}_{ \widetilde{ \mathbf{ \Omega }} } \right\}^j \mathbf{S}^{\prime}
\end{align} 
Therefore, we can conclude that $\widetilde{ \mathbf{\Gamma} }^{-1} \equiv \mathbf{S} \mathbf{D}_{ \widetilde{\lambda} } \mathbf{S}^{\prime}$ where we denote with $\mathbf{D}_{ \widetilde{\lambda} }$ the following matrix 
\begin{align}
\mathbf{D}_{ \widetilde{\lambda} } := diag \left( \overset{\infty}{\underset{ j = 0 }{\sum}} \left( \lambda_1^{ \widetilde{ \Omega} } \right)^j , ..., \overset{\infty}{\underset{ j = 0 }{\sum}} \left( \lambda_N^{ \widetilde{ \Omega} } \right)^j  \right).
\end{align}
That is, $\mathbf{D}_{ \widetilde{\lambda} } \in \mathbb{R}^{ N \times N}$ is a diagonal matrix with diagonal elements $\overset{\infty}{\underset{ j = 0 }{\sum}} \left( \lambda_k^{ \widetilde{ \Omega} } \right)^j$, where $\lambda_k^{ \widetilde{ \Omega} }$ is an eigenvalue of the matrix $\widetilde{ \mathbf{ \Omega }}$. The particular result holds if and only if these eigenvalues are bounded, such that $0 < \lambda_k^{ \widetilde{ \Omega} } < 1$ for all $k \in \left\{ 1,..., N \right\}$ where $N$ the number of stocks. 

\newpage 

\begin{remark}
From the last expression, we basically proved that an eigenvalue of the matrix $\widetilde{\mathbf{\Gamma}}^{-1}$ is related to an eigenvalue of the matrix $\left( \mathbf{I}_N - \widetilde{ \mathbf{ \Omega }} \right)^{-1}$ in the following way
\begin{align}
\label{eigen.gamma.inverse}
\lambda^{ \widetilde{ \mathbf{\Gamma} }^{-1} }_k \equiv  \overset{\infty}{\underset{ j = 0 }{\sum}} \left( \lambda_k^{ \widetilde{ \Omega} } \right)^j = \frac{1}{1 -  \lambda_k^{ \widetilde{ \Omega} } }
\end{align}
The absolute convergence of the infinite series sum holds if and only if the eigenvalues of the matrix $\widetilde{ \mathbf{ \Omega }}$ are bounded between 0 and 1 (see, Assumption \ref{assumption3}). Then, we can conclude that the elements of $\mathbf{D}_{ \widetilde{\lambda} }$ are given by $1 \big/ \left( 1 - \lambda_k^{ \widetilde{ \Omega} } \right)$ for $k \in \left\{ 1,\ldots,N \right\}$. 
\end{remark}

By employing the expression we proved above, $\widetilde{ \mathbf{\Gamma} }^{-1} \equiv \mathbf{S} \mathbf{D}_{ \widetilde{\lambda} } \mathbf{S}^{\prime}$, we can now express the optimal portfolio weights of the minimum variance portfolio problem as below
\begin{align}
\label{expression.opt}
\mathbf{w}^{*} 
= 
\frac{ \mathbf{S} \mathbf{D}_{ \widetilde{\lambda} } \mathbf{S}^{\prime} \mathbf{1}}{\mathbf{1}^{\prime} \mathbf{S} \mathbf{D}_{\widetilde{\lambda} } \mathbf{S}^{\prime} \mathbf{1}}.
\end{align}

Furthermore, by expanding the closed-form solution given by \eqref{expression.opt} we can express each optimal portfolio weight $w_i^*$ as a function of the eigenvector centrality $\widetilde{v}_i$. The trick for obtaining an expression with the eigenvector centrality is to decompose the SVD into two terms, such as the first term corresponds to the largest eigenvalue and the second term corresponding to the remaining $(N-1)$ eigenvalues given by expression \eqref{eigen.gamma.inverse}. By simple matrix algebra, we obtain the following expression
\begin{align}
\label{the.weight}
w_i^{*} 
= \frac{ \displaystyle \frac{ \widetilde{v}_i }{ 1 - \lambda_{(1)}^{ \widetilde{ \Omega} } } \left( \overset{N}{\underset{j=1}{\sum}} \widetilde{v}_j \right) + \overset{N}{\underset{ k=2 }{ \sum} } \frac{ s_{ik} }{1-\lambda_k^{ \widetilde{ \Omega} } } \left( \overset{N}{\underset{j=1}{\sum} } s_{jk} \right) }{ \displaystyle \frac{1}{ 1 - \lambda_{(1)}^{ \widetilde{ \Omega} } } \left(\overset{N}{\underset{j=1}{\sum}} \widetilde{v}_j \right)^2 + \overset{N}{\underset{k=2}{\sum}} \frac{1}{1-\lambda_k^{ \widetilde{ \Omega} } } \left(\overset{N}{\underset{j=1}{\sum}} s_{jk} \right)^2},
\end{align}
where $\lambda_{(1)}^{ \widetilde{ \Omega} }$ the largest eigenvalue of $\widetilde{ \mathbf{ \Omega } }$ and $\widetilde{v}_i$ the corresponding centrality measure based on the adjacency matrix  $\widetilde{ \mathbf{ \Omega } }$ where $s_{ik}$ denotes the $(i,k)-$element of the matrix $\mathbf{S}$. 
\end{proof}

\medskip

\begin{remark}
The diagonal matrix $\mathbf{D}_{ \widetilde{\lambda} }$ contains the leading eigenvalue of the symmetrized risk matrix $\widetilde{ \mathbf{\Gamma} }$. The particular leading eigenvalue which is an important quantity within our framework might suffer from larger bias than the leading eigenvalue of $\mathbf{\Gamma}$ which is the asymmetric matrix (see, \cite{chen2021asymmetry} for a discussion about this aspect). However, the symmetrization approach we follow avoids the need for complex estimation procedure when obtaining sample estimates of other quantities such as the inverse of the risk matrix, $\widetilde{ \mathbf{\Gamma} }^{-1}$. We aim to investigate  related aspects to bias of the leading eigenvalue in a further study. 
\end{remark}

\newpage 

\noindent \textbf{Proof of Theorem \ref{theorem2}:}

\begin{proof}
Applying the decomposition given by \eqref{lambda1b} to the adjacency matrix $\widetilde{ \mathbf{ \Omega } }$ we obtain that 
\begin{equation}
\label{frob}
\lambda_{ (1) }^{ \widetilde{ \Omega} }  
=
\overset{N}{\underset{i=1}{\sum}} s_{i1}^2 \left( 1 - \gamma_i \right) + \overset{N}{\underset{i=1}{\sum}} \overset{N}{\underset{\underset{j \neq i}{j=1}}{\sum}} s_{i1} s_{j1} \widetilde{ \Omega}_{ij}. 
\end{equation}
where $\lambda_{ (1) }^{ \widetilde{ \Omega} }$ is the Frobenius eigenvalue of the matrix $\widetilde{ \mathbf{ \Omega } }$. Notice that expression \eqref{frob} decomposes the largest eigenvalue into two terms, with the first term showing clearly the dependence of the largest eigenvalue to the tail risk measure of the main diagonal of the matrix. 

Replacing by the eigenvector centrality measures in \eqref{central1}, we obtain that
\begin{equation}
\lambda_{ (1) }^{ \widetilde{ \Omega} } 
= 
\overset{n}{\underset{i=1}{\sum}} \widetilde{ v }_{i}^{2} \left( 1 - \gamma_i \right)  + \overset{N}{\underset{i=1}{\sum}} \overset{N}{\underset{\underset{j \neq i}{j=1}}{\sum}} \widetilde{ v }_{i} \widetilde{ v }_{j} \widetilde{ \Omega }_{ij}. 
\end{equation}

Then, the first derivative with respect to the eigenvector centrality measure $\widetilde{ v }_i$ is 
\begin{align}
\frac{ \partial \lambda_{ (1) }^{ \widetilde{ \Omega } } }{  \partial \widetilde{ v }_i } = 2 \widetilde{ v }_i \left( 1 - \gamma_i \right)  + \overset{N}{ \underset{\underset{j \neq i}{j=1}}{\sum}} \widetilde{ \Omega } \widetilde{ v }_{j}. 
\end{align}

By definition of eigenvector centrality in \eqref{central1}, we obtain that
\begin{align}
\left( 1 - \gamma_i \right) \widetilde{ v }_i + \overset{N}{\underset{\underset{j \neq i}{j=1}}{\sum}} \widetilde{ \Omega }_{ij} \widetilde{ v }_{j} = \lambda_{ (1) }^{ \widetilde{ \Omega } } \widetilde{ v }_i \ \ 
\end{align}

such that 
\begin{align}
\frac{ \partial \lambda_{(1)}^{ \widetilde{ \Omega } } }{  \partial \widetilde{v}_i }  = \big( 1 - \gamma_i + \lambda_{(1)}^{ \widetilde{ \Omega } }  \big) \widetilde{ v }_i 
\end{align}

To prove the result of Theorem \ref{theorem2} we need to show that the numerator of the first derivative of the quantity $w_i^*$ as a function of $\widetilde{ v }_i$ is positive. Furthermore, notice that due to the construction of the original adjacency matrix $\mathbf{ \Omega }$ as a weighted adjacency matrix, we avoid the need to impose additional assumptions regarding the sparsity of the matrix. Therefore, the eigenvalues and eigenvectors of both  $\mathbf{ \Omega }$ and $\widetilde{ \mathbf{ \Omega } }$ matrices are well-behaved. 

\begin{remark}
Notice that for the result of Theorem \ref{theorem2} to hold an important assumption is that all the elements of the centrality vector $\widetilde{v}_i$ are positive in sign. In practise, in empirical applications this might not always hold, if one uses the eigenvector which corresponds to the largest eigenvalue as the centrality measure. Therefore, by using the \texttt{graph} package in R we can obtain a positive eigenvector centrality vector based on an adjacency matrix. 
\end{remark}

\newpage 

These conditions are given as following
\begin{itemize}
\item[(i)]   $\overset{N}{\underset{j=1}{\sum}} s_{jk}  - 2 s_{ik}>0$ for $i,k \in \left\{ 1,\dots,N \right\}$;
\item[(ii)]  $\overset{N}{\underset{j=1}{\sum}} \widetilde{v}_{j} > 1 $; 
\item[(iii)] $\widetilde{v}_{i} \left(\overset{N}{\underset{j=1}{\sum}} s_{jk} \right) - \displaystyle \frac{ s_{jk} }{1 - \lambda_{ (1) }^{ \widetilde{ \Omega} } }>0 $ for $i,k \in  \left\{ 1,\dots,N \right\}$. 
\end{itemize}

Thus, Assumption \ref{assumption4} guarantees that these conditions are satisfied. Notice that these conditions are sufficient but not necessary to obtain a positive relationship between stock centrality and the optimal allocation to the risky asset, under the assumption of short selling. 
\end{proof}

\bigskip

\noindent \textbf{Proof of Proposition \ref{proposition8}:}

\begin{proof}
To prove this result we first show that $s_k=z_k$ for $k \in \left\{ 1,\ldots,N \right\}$, with $s_k$ and $z_k$ the eigenvectors associated to the adjacency matrices $\widetilde{ \mathbf{\Omega}} = - \left( \mathbf{B} + \mathbf{\Omega} \right)$ and $\mathbf{\Omega}$, respectively. This is equivalent to showing that the vectors $z_k$ are also eigenvectors of $\widetilde{ \mathbf{\Omega}}$, implying that $\widetilde{ \mathbf{\Omega}} z_k= \lambda_k^{ \widetilde{ \Omega} } z_k$ is satisfied. Thus, we focus on proving that this condition holds. We have that $diag ( \widetilde{\mathbf{\Gamma}} ) = \gamma \otimes \mathbf{I}_N$ when the VaR measures $\gamma_i$ are common across assets. Then, the adjacency matrix is written as $\widetilde{ \mathbf{\Omega}} = \left( 1 - \gamma \right) \mathbf{I}_N - \mathbf{ \Omega }$. Moreover by construction, $\mathbf{ \Omega } z_k= \lambda_k^{ \Omega } z_k $, such that
\begin{equation}
\widetilde{ \mathbf{\Omega}} z_k =  \left( 1 - \gamma \right) \mathbf{I}_N z_k - \mathbf{\Omega } z_k = \left( 1 - \gamma \right) z_k - \lambda_k^{ \Omega } z_k = \lambda_{(1)}^{ \Omega } z_k.
\end{equation}

\medskip

Therefore, it follows that the eigenvectors of $\mathbf{ \Omega }$ are also the eigenvectors of $\widetilde{ \mathbf{\Omega}}$. Due to this  result we also have that the eigenvector centrality measures associated to the adjacency matrices $\mathbf{\Omega}$ and  $\widetilde{ \mathbf{\Omega}}$ satisfy that $v_k = \widetilde{v}_k$ for $k \in \left\{ 1,\ldots,N \right\}$. Notice that the eigenvalues satisfy that $\lambda_k^{ \widetilde{ \Omega} } = \left( 1 - \gamma - \lambda_k^{ \Omega } \right)$ for $k \in \left\{ 1,\ldots,N \right\}$. Then, $1-\lambda_k^{ \widetilde{ \Omega} } = \gamma + \lambda_k^{ \Omega }$, and replacing the eigenvectors and eigenvalues in \eqref{weight6}, we obtain the expression for the optimal asset allocation
\begin{align}
\mathbf{w}_i^{*} 
= \frac{ \displaystyle \frac{ v_{i} }{ \gamma + \lambda_{(1)}^{ \Omega } } \left(\overset{N}{\underset{j=1}{\sum}} v_{j} \right) + \overset{N}{\underset{k=2}{\sum}} \frac{z_{ik}}{ \gamma + \lambda_k^{ \Omega }  } \left(\overset{N}{\underset{j=1}{\sum}} z_{jk} \right)}{ \displaystyle  \frac{1}{\gamma + \lambda_{(1)}^{ \Omega }  } \left(\overset{N}{\underset{j=1}{\sum}} v_{j} \right)^2 + \overset{N}{\underset{k=2}{\sum}} \frac{1}{\gamma + \lambda_k^{ \Omega }  } \left(\overset{N}{\underset{j=1}{\sum}} z_{jk} \right)^2},
\end{align}
such that $\displaystyle  \frac{\partial w_i^*}{\partial v_i} = \frac{ \partial w_i^{*} }{ \partial \widetilde{v}_i }$, and, hence, applying Theorem \ref{theorem2}, we find that the optimal allocation to asset $i$ is increasing on stock centrality $v_i$, proving the statement of Proposition  \ref{proposition8}.
\end{proof}

\newpage 

\section{Estimation Results}
\label{AppendixB}

\begin{table}[h!]
  \centering
  \caption{\\ Variable Definitions}
    \begin{tabular}{ll}
    \hline
    \textbf{Firm Level} & Financial Variable Definition \\
    \hline
    Leverage  & Total Assets / Total Equity \\
    \hline
    Maturity Mismatch  & Short term Debt / Total liabilities \\
    \hline
    Size  & Log of Total Book Equity \\
    \hline
    Market-to-book  & Market Value of Total Equity / Book Value of Total Equity \\
    \hline
          &  \\
          \hline
    \textbf{Macroeconomic} & Financial Variable Definition \\
    \hline
    Variable 1 & the implied volatility index (VIX) \\
    \hline
    \multirow{2}[0]{*}{Variable 2} & the short term liquidity spread calculated as the difference between the  \\
          & three-month repo rate and the three-month bill rate  \\
          \hline
    Variable 3 & the changes in the three-month Treasury bill rate. \\
    \hline
    \multirow{2}[0]{*}{Variable 4} & the changes in the slope of the yield curve corresponding to the yield spread \\
          & between the ten year Treasury rate and the three-month bill rate from FRB \\
          \hline
    Variable 5 & the changes in the credit spread between BAA rated bonds and \\
          & the Treasury rate. \\
          \hline
    Variable 6 & the weekly S\&P500 index returns. \\
    \hline
    Variable 7 & the weekly Dow Jones US Real Estate index returns. \\
    \hline
    \end{tabular}%
  \label{table 1}%
\end{table}%

\begin{small}
Table \ref{table 1} gives a summary of the financial and macroeconomic variables included in the dataset of \cite{hardle2016tenet} which we utilize for our empirical application. Notice that we assume  stationarity which avoids existence of structural breaks in certain macroeconomic variables such as the VIX\footnote{Note that the VIX is considered a robust financial variable which can capture financial conditions and time-varying volatility effects across the US financial markets. A detailed study on risk premia and stock characteristics can be found in \cite{martin2019expected}.}.   
\end{small}

\begin{table}[h!]
  \centering
  \caption{\\ Main Statistics }
    \begin{tabular}{cccccccccc}
     \hline
          & \textbf{$T$} & Mean & SD & SR & Min & Q1 & Median & Q3 & Max \\
     \hline
    \textbf{X} & 63    & 0.0032 & 0.0041 & 0.7780 & -0.0732 & -0.0142 & 0.0042 & 0.0212 & 0.0967 \\
    \textbf{Y} & 63    & 0.0038 & 0.0033 & 1.1476 & -0.0621 & -0.0099 & 0.0047 & 0.0214 & 0.0866 \\
    \textbf{Z} & 63    & 0.0036 & 0.0036 & 0.9867 & -0.0684 & -0.0112 & 0.0036 & 0.0223 & 0.0896 \\
     \hline
    \end{tabular}%
  \label{table2}%
\end{table}%

\begin{small}
Table \ref{table2} presents the summary statistics for the distribution of portfolio returns for the stationary macroeconomic variables based on taking first differences on the full sample. 
\end{small}

\medskip

We denote with $X$ to denote the out-of-sample distribution of portfolio returns for the network topology with 25 less central assets removed (i.e., high interconnectedness); and $Y$ represents the out-of-sample distribution of portfolio returns for the network topology with 25 most central assets removed (i.e., low interconnectedness). The results presented on Table \ref{table2} corresponds to the optimal portfolio allocation problem (GMVP) using the proposed risk matrix (and corresponding adjacency matrix) under the assumption of stationary regressors. 

\newpage

We ensure that stationarity holds by taking the first difference of the macro variables  within each out-of-sample period separately. Moreover, the vector of optimal weights is reestimated in each iteration following the dynamic portfolio optimization framework. The particular statistics which correspond to the moments of the distributions of portfolio returns, indicate that the less interconnected is the network topology the higher is the Sharpe ratio. More precisely, these estimates correspond to $N = 100$ nodes, for a rolling window of size $n = 250$, resulting to an out-of-sample period of $T = 63$ time observations. Furthermore, in each out-of-sample period we exclude 25 more central assets to obtain a network topology of low inteconnectedness ($Y$ distribution of portfolio returns) and similarly we apply the same forecasting scheme but excluding 25 less central assets to obtain a network topology of high inteconnectedness ($X$ distribution of portfolio returns). Additionally, the estimates that correspond to the full network of each rolling window are obtained for comparability purposes. Lastly, the aforementioned methodology is implemented for estimation windows of different window size to evaluate its effect to the robustness of the estimates. The corresponding moments from the out-of-sample distributions of portfolio returns are reported on Table \ref{table2}.  

\begin{table}[h!]
  \centering
  \caption{\\ Out-of-sample Portfolio Returns Distribution Estimates }
     \begin{tabular}{cccccccccc}
    \hline
          & \textbf{$T$} & Mean & SD & SR & Min & Q1 & Median & Q3 & Max \\
    \hline
    \textbf{X} & 63    & 0.0032 & 0.0227 & 0.1422 & -0.0576 & -0.0101 & 0.0034 & 0.0177 & 0.0698 \\
    \textbf{Y} & 63    & 0.0038 & 0.0207 & 0.1824 & -0.0486 & -0.0082 & 0.0051 & 0.0163 & 0.0614 \\
    \textbf{Z} & 63    & 0.0036 & 0.0196 & 0.1856 & -0.0446 & -0.0088 & 0.0052 & 0.0154 & 0.0559 \\
    \hline
    \end{tabular}%
  \label{tableA}%
\end{table}%

\begin{small}
Table \ref{tableA} presents main moment estimates for the distribution of portfolio returns for $10\%$ exclusion of central nodes based on stationary regressors within each out-of-sample period. 
\end{small}

\begin{table}[h!]
  \centering
  \caption{\\ Out-of-sample Portfolio Returns Distribution Estimates }
     \begin{tabular}{cccccccccc}
      \hline
          & \textbf{$T$} & Mean & SD & SR & Min & Q1 & Median & Q3 & Max \\
      \hline
    \textbf{X} & 63    & 0.0030 & 0.0270 & 0.1108 & -0.0599 & -0.0125 & 0.0026 & 0.0214 & 0.0830 \\
    \textbf{Y} & 63    & 0.0038 & 0.0214 & 0.1767 & -0.0506 & -0.0096 & 0.0057 & 0.0164 & 0.0634 \\
    \textbf{Z} & 63    & 0.0036 & 0.0196 & 0.1856 & -0.0446 & -0.0088 & 0.0052 & 0.0154 & 0.0559 \\
     \hline
    \end{tabular}%
  \label{tableB}%
\end{table}%

\begin{small}
Table \ref{tableB} presents main moment estimates for the distribution of portfolio returns for $25 \%$ exclusion of central nodes based on stationary regressors within each out-of-sample period.
\end{small}

\newpage 

\section{Additional Results}
\label{AppendixD}

Using the Leibniz' rule, the $n-$th derivative of both sides of $A v( \lambda ) = \lambda v( \lambda )$, with respect to $\lambda$ is given by the expression below
\begin{align}
A \frac{d^n v( \lambda ) }{ d \lambda^n } = \sum_{ k = 0 }^n  \binom{n}{k} \frac{ d^k }{ d \lambda^k  } ( \lambda ) = \lambda \frac{ d^n v( \lambda ) }{ d \lambda^n  } + n \frac{ d^{n-1} v( \lambda ) }{  d \lambda^{n-1} }
\end{align} 
Thus, for $n \leq 1$ we obtain that 
\begin{align}
\left( A - \lambda I \right) \frac{ d^{n} v( \lambda ) }{  d \lambda^{n} }  = n \frac{ d^{n-1} v( \lambda ) }{  d \lambda^{n-1} }
\end{align}
Notice that further details regarding the theoretical proofs on differentiating eigenvalues and eigenvectors can be found in the paper of \cite{magnus1985differentiating} (see, also \cite{phillips1982simple}). Moreover, further background on related mathematical theorems can be found in \cite{van2010graph}. In terms of the estimation procedure for the spectral radius of a positive definite matrix (see, Theorem \ref{PerronFrobenius}), a relevant procedure is proposed by \cite{ibragimov2001method}. 

\medskip

\begin{theorem}
\label{PerronFrobenius}
\textbf{(Perron Frobenius)} \citep{van2010graph} An irreducible non-negative $n \times n$ matrix $A$ always has a real, positive eigenvalue $\lambda_1 = \lambda_{ \text{max} } \left( A \right)$ and the modules of any other eigenvalue does not exceed $\lambda_{ \text{max} }$, that is, $| \lambda_k \left( A \right) | \leq \lambda_{ \text{max} }$ for $k = 2,...,n$. Moreover, $\lambda_1$ is a simple zero of the characteristic polynomial det$\left( A - \lambda I \right)$. The eigenvector belonging to $\lambda_1$ has positive components. 
\end{theorem}

\begin{remark}
Notice that by the Perron Frobenius theorem, the spectral radius of a matrix $A$ is its nonnegative eigenvalue, to which a positive eigenvector corresponds. Furthermore, the spectrum of a matrix $A$ is a sorted sequence of eigenvalues. For instance, we can denote the spectrum of a matrix $A$ in terms of a descending sequence such that $\lambda^A_1 \geq \lambda^A_2 \geq ... \geq \lambda^A_n$. 
\end{remark}

\subsection{Properties of Perturbation Matrix}

Another useful result is presented by Corollary \ref{corollary1} below.  In particular, we show that the eigenvector centrality measure for a node $i$ is the same for all adjacency matrices obtained from parallel shifts of $\mathbf{\Omega}$. Therefore, we consider the following affine transformation 
\begin{align}
\label{OmegaMatrix}
\mathbf{\Omega}^{\star} := \mathbf{\Omega} + \eta \hspace{0.2ex} \mathbf{I}_N
\end{align}
with $\mathbf{I}_N$ the $N \times N$ identity matrix and $\eta \in \mathbb{R}$ some positive constant.

\medskip 

\begin{corollary} 
\label{corollary2}
The centrality measure $v_i$ for stock $i$ with adjacency matrix $\mathbf{ \Omega }^{\star} := \mathbf{\Omega} + \eta \hspace{0.2ex} \mathbf{I}_N$ is equivalent to the corresponding centrality measure for the adjacency matrix $\mathbf{\Omega}$, where $\eta$ represents the perturbation parameter.
\end{corollary}

\newpage 

\begin{proof}
By definition, the centrality of stock $i$ given by $v_i$ is defined by expression \eqref{central0}
\begin{align} 
\label{central}
v_i^{\star} = \left( \lambda^{ \Omega^{\star} }_{(1)} \right)^{-1} \overset{N}{\underset{j=1}{\sum}} \Omega^{\star}_{ij} \hspace{0.2ex} v^{\star}_j,
\end{align}
where $\lambda^{ \Omega^{\star} }_{(1)} := \max \left( \lambda^{ \Omega^{\star} }_1,\ldots, \lambda^{ \Omega^{\star} }_N \right)$ is the largest eigenvalue of the adjacency matrix $\mathbf{ \Omega }^{\star}$. 

This result follows immediately by considering that expressions \eqref{central0} and \eqref{central} are equal for the centrality measures based on the adjacency matrices $\mathbf{ \Omega }$ and $\mathbf{ \Omega }^{\star}$ respectively. To show this, note from \eqref{central} we obtain that 
\begin{align}
\lambda^{ \Omega^{\star} }_{(1)} v^{\star}_i = \eta + \overset{N}{ \underset{\underset{i \neq j}{j=1}}{\sum}} \Omega^{\star}_{ij} \hspace{0.2ex} v^{\star}_j. 
\end{align}
Furthermore, the eigenvalues of the adjacency matrices $\mathbf{\Omega}^{\star}$ and $\mathbf{ \Omega }$ are related such that $\lambda^{ \Omega^{\star} }_i =\lambda_i^{ \Omega } + \eta$, for all $i \in \left\{ 1, \ldots, N \right\}$. This is an important result which provides an equivalence relation of the corresponding set of eigenvalues, which is similar to the affine transformation of the adjacency matrix $\mathbf{ \Omega }$. To see this, we denote with $ \mathbf{Z}_{\Omega^{\star}} = \left[ \mathbf{z}^{\Omega^{\star}}_1,\ldots, \mathbf{z}^{\Omega^{\star}}_N \right]$ the matrix of eigenvectors of $\mathbf{\Omega}^{\star}$ and similarly we denote with $\mathbf{Z}_{\Omega} = \left[ \mathbf{z}^{ \Omega }_1,\ldots, \mathbf{z}^{ \Omega }_N \right]$ be the eigenvector matrix of the adjacency matrix $\mathbf{\Omega}$. Then, by the definition of the eigenvalue problem we have that $\lambda^{\Omega^{\star}}_i \mathbf{ z}^{\Omega^{\star}}_i = \mathbf{\Omega} \mathbf{z}^{\Omega}_i$. Therefore, by substituting the expression for $\mathbf{\Omega}^{\star}$, we obtain that
\begin{align}
\lambda^{\Omega^{\star}}_i \mathbf{ z}^{\Omega^{\star}}_i  = \left( \mathbf{\Omega} + \eta \hspace{0.2ex} \mathbf{I}_N \right) z^{\Omega^{\star}}_i = (\lambda_i^{\Omega } + \eta ) z^{\Omega^{\star}}_i
\end{align}

which gives that $\lambda^{\Omega^{\star}}_i = \lambda^{ \Omega }_i + \eta$, relating the eigenvalues of the two adjacency matrices. 

Notice that by construction the off-diagonal elements of both adjacency matrices are also equal such that $\Omega^{\star}_{ij} = \Omega_{ij}$ for all $i \neq j$. Therefore, it follows that 
\begin{align}
\lambda^{\Omega}_{i} v_i = \overset{N}{\underset{i \neq j}{\underset{j=1}{\sum}}} \Omega_{ij} \hspace{0.2ex} v_j
\end{align} 
which proves that condition \eqref{central} is satisfied.
\end{proof}

\newpage

\bibliographystyle{apalike}
\small{
\bibliography{myreferences1}}

\newpage

\setlength{\parindent}{.0cm} \setlength{\parskip}{.1cm}
\end{document}